\documentclass[letterpaper, 10 pt, journal]{ieeeconf}
\usepackage{cite}
\usepackage{amsmath,amssymb,amsfonts}
\usepackage{algorithmic}
\usepackage{graphicx}
\usepackage{textcomp}
\usepackage{xcolor}
\usepackage{array, booktabs}
\let\labelindent\relax
\usepackage{enumitem}
\usepackage{dsfont}
\usepackage{siunitx}
\usepackage{float}
\usepackage{placeins}
\usepackage{footmisc}
\usepackage{xurl}
\usepackage{hyperref}

\def\BibTeX{{\rm B\kern-.05em{\sc i\kern-.025em b}\kern-.08em
    T\kern-.1667em\lower.7ex\hbox{E}\kern-.125emX}}

\urldef{\vidurl}\url{https://tinyurl.com/hierracingvideos}
\urldef{\codeurl}\url{https://www.github.com/ribsthakkar/HierarchicalKarting}

\makeatletter
\newcommand\extralabel[2]{{\edef\@currentlabel{\@currentlabel#2}\label{#1}}}
\makeatother
\IEEEoverridecommandlockouts

\begin{document}
\author{
    \authorblockN{Rishabh Saumil Thakkar, Aryaman Singh Samyal}, David Fridovich-Keil, Zhe Xu, Ufuk Topcu
    \thanks{
    R. S. Thakkar is with the Oden Institute of Computational Engineering and Sciences and A. S. Samyal, D. Fridovich-Keil, and U. Topcu are with the Department of Aerospace Engineering and Engineering Mechanics at The University of Texas at Austin (email: \{\mbox{rishabh.thakkar}, \mbox{aryamansinghsamyal}, \mbox{dfk}, \mbox{utopcu}\}\mbox{@utexas.edu}). \newline 
    \indent\indent Z. Xu is with the School for Engineering of Matter, Transport, and Energy at Arizona State University (email: \mbox{xzhe1@asu.edu}).
    }
}
\title{Hierarchical Control for Cooperative Teams \\ in Competitive Autonomous Racing}
\maketitle
\begin{abstract}
We investigate the problem of autonomous racing among teams of cooperative agents that are subject to realistic racing rules. Our work extends previous research on hierarchical control in head-to-head autonomous racing by considering a generalized version of the problem while maintaining the two-level hierarchical control structure. A high-level tactical planner constructs a discrete game that encodes the complex rules using simplified dynamics to produce a sequence of target waypoints. The low-level path planner uses these waypoints as a reference trajectory and computes high-resolution control inputs by solving a simplified formulation of a racing game with a simplified representation of the realistic racing rules. We explore two approaches for the low-level path planner: training a multi-agent reinforcement learning (MARL) policy and solving a linear-quadratic Nash game (LQNG) approximation. We evaluate our controllers on simple and complex tracks against three baselines: an end-to-end MARL controller, a MARL controller tracking a fixed racing line, and an LQNG controller tracking a fixed racing line. Quantitative results show our hierarchical methods outperform the baselines in terms of race wins, overall team performance, and compliance with the rules. Qualitatively, we observe the hierarchical controllers mimic actions performed by expert human drivers such as coordinated overtaking, defending against multiple opponents, and long-term planning for delayed advantages.
\end{abstract}


\section{Introduction}
Autonomous driving has seen a rapid growth of research in academia and industry \cite{adlit}. While most of these efforts focus on day-to-day driving, there is growing interest in autonomous racing \cite{litreview}. Many advances in commercial automobiles have originated from projects invented for use in motorsports such as disc brakes, rear-view mirrors, and sequential gearboxes \cite{racingadvances}. The same principle can apply when designing self-driving controllers because racing provides a platform to develop these controllers to be high-performance, robust, and safe in challenging scenarios.

Successful human drivers are required to outperform opponents and adhere to the rules of racing. These objectives are effectively at odds with each other, but the best racers can achieve both. Prior approaches in autonomous racing usually over-simplify the latter by only considering basic collision avoidance \cite{Wang2019, Wang2021, Li2021, He2021}. 

In reality, racing rules often involve discrete logic and complex nuances \cite{racingrules}. For example, a driver may not change lanes more than a fixed number of times when traveling along a straight section of the track. While it is relatively straightforward to describe this rule in text, it is challenging to encode it in a mathematical formulation that can be solved by existing methods for real-time control. Methods such as model predictive control have to compromise by either using short planning horizons or simply ignoring these constraints \cite{Wang2019, Wang2021}. 

\begin{figure}
  \centering
  \includegraphics[width=0.48\textwidth]{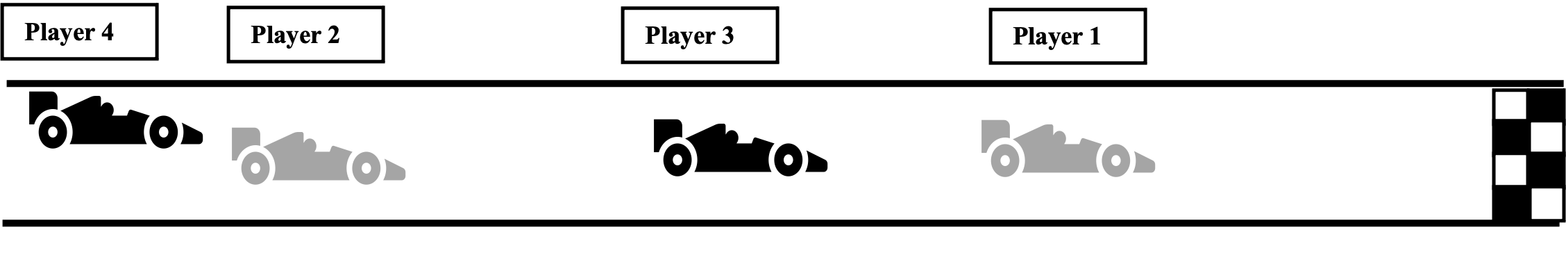}
  \caption[Motivating example for team-based racing] {Because players have incentive to finish ahead as a team in addition to improving their own finishing position, Player 3's strategy is unclear. Is there enough time to try to pass Player 1 before the finish line? Otherwise, should it consider slowing down on purpose to try help Player 4 pass Player 3 at the risk of being overtaken itself or simply maintain position? }
  \label{fig:team_motivating}
\end{figure}

In addition, real-life racing also involves an aspect of teamwork where drivers have one or more teammates, and there is an additional objective of collectively finishing ahead of other teams. Therefore, drivers are required to race with a combination of cooperative and competitive objectives in mind while continuing to adhere to complex safety and fairness rules. In such scenarios, determining the best strategy is not trivial and requires drivers to evaluate the long-term impacts of their choices. Consider the example in Figure \ref{fig:team_motivating}. Player 1 and Player 2 are on one team, and Player 3 and Player 4 are on another team. Player 1 is clearly first and almost at the finish line, so it is unlikely that Player 3, who is in second, can catch him before the finish line. On the other hand, Player 4 is in last, but it is close to Player 2 in third. Player 3 now has three high-level choices to consider:
\begin{enumerate}
    \item Try to overtake Player 1 before the finish line.
    \item Maintain its position to the finish line.
    \item Purposely slow down to block Player 2 and improve the chances of Player 4 overtaking Player 2 at the risk of being overtaken by Player 2.
\end{enumerate}
If all players are racing independently, choice 1 would likely be the most reasonable because that is only possibility of any payoff. However, in the cooperative team setting, because there is an incentive to finish higher overall as a team, Player 3 must consider the payoffs and risks associated with all three choices. These factors are not obvious to evaluate because the implications of the choices are not immediately observed, and it is usually challenging to switch from one choice to another. For example, committing to the choice 3 means that Player 3 cannot realistically change its mind and switch to choice 1 if it realizes the risk is too high. 

This paper builds on previous work on hierarchical control in head-to-head racing \cite{thakkarprior}. The contributions of this paper begin with developing a generalized version of a the racing game with realistic safety rules introduced in the previous paper. Our updated formulation allows for $N$ players organized into teams, and includes a mixed objective that takes into account both individual and team performance.

\begin{figure}
  \centering
  \includegraphics[width=0.48\textwidth]{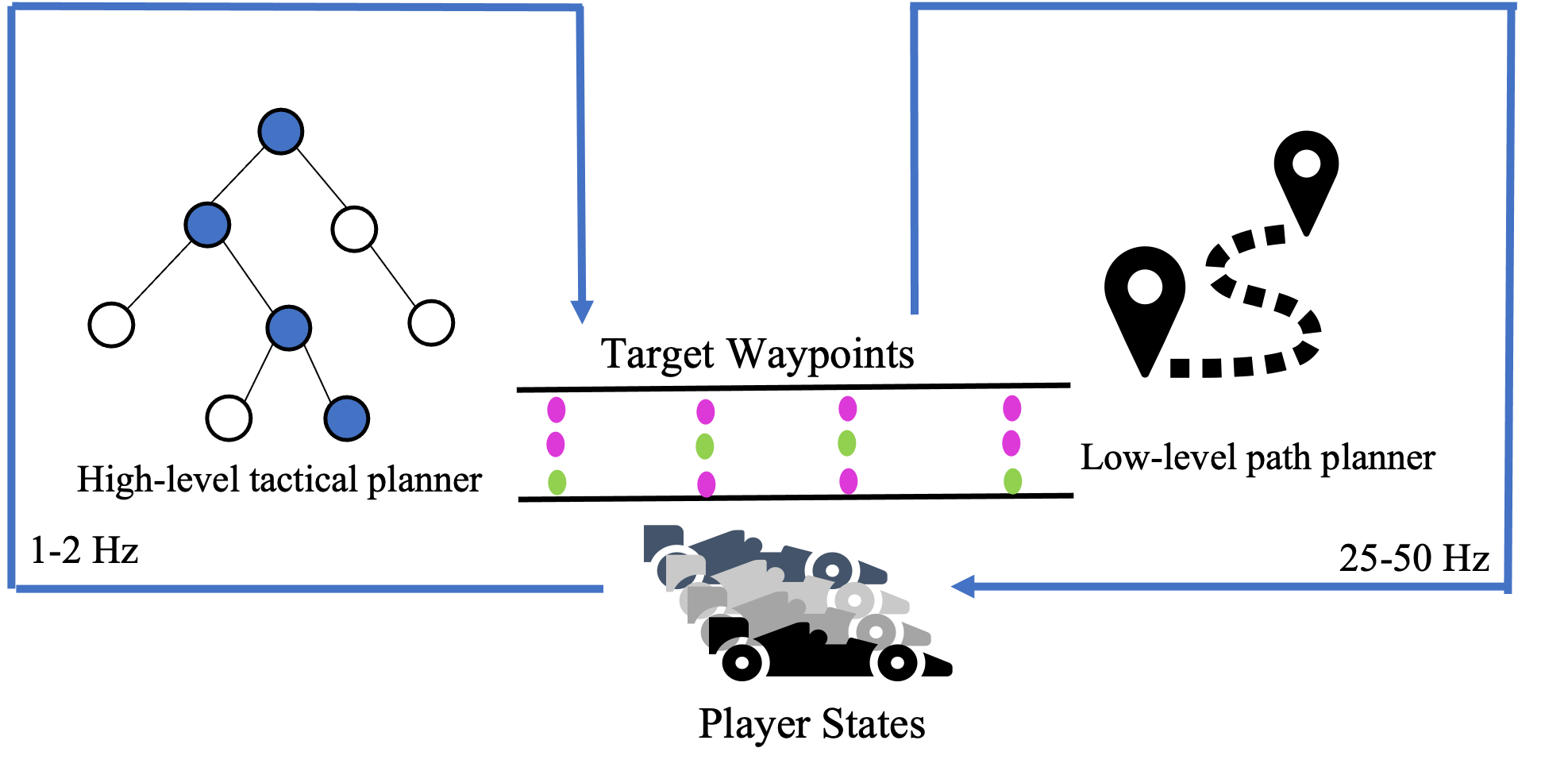}
  \caption{Two-level planning architecture of the proposed racing controller.}
  \label{fig:control_arch}
\end{figure}

We then design a two-level hierarchical controller to tackle this complex problem. The controller considers both competitive and cooperative objectives and enables us to consider the safety rules in real-time. The two levels of the controller consist of a high-level tactical planner and a low-level path planner. The high-level planner creates a discrete approximation of the general formulation, which makes it easy to model the discrete nature of the safety rules. The output of the high-level planner is a series of target waypoints. Given these waypoints, the low-level path planner solves a simplified continuous state/action dynamic game to produce control inputs that aim to reach the waypoints as closely as possible, help teammates pass as many waypoints as possible, and prevent opposing teams from passing waypoints while taking into account a simplified version of the safety rules. The control architecture is shown in Figure \ref{fig:control_arch}.

Finally, we show that our hierarchical planning model outperforms other common approaches in a high-fidelity simulator in terms of maximizing both cooperative and competitive measures of performance and adherence to the safety rules. To our knowledge, this is the first paper to study teamwork in the context of autonomous racing.  And while we develop our controller in the context of a racing game, the structure of our approach makes it possible to reason about long-term optimal choices in more general game-theoretic settings with complex objectives, constraints involving temporal logic, and both continuous and discrete dynamics. 
This makes it possible to apply our method to other adversarial settings with similar properties, such as financial systems, power systems, or air traffic control \cite{Xu2021, Tomlin1996}.

\section{Prior Work}
Because multi-agent racing is inherently a more complex problem than single-agent lap time optimal control, most prior work in autonomous racing is focused on single-agent lap time optimization, with fewer and more recent developments in multi-agent racing.

Single-agent racing approaches include both optimization and learning-based methods. One study uses Monte Carlo tree search to estimate where to position the car around various shaped tracks to define an optimal trajectory \cite{Hou2016}. Another paper proposes a method that computes an optimal trajectory offline and uses a model predictive control (MPC) algorithm to track the optimized trajectory online \cite{Vazquez2020}. Similarly, the Stahl et al. \cite{Stahl2019_2} also perform calculations offline by creating a graph representation of the track to compute a target path and use spline interpolation for online path generation in an environment with static obstacles. In the category of learning-based approaches, Kabzan et. al \cite{Kabzan2019} use online learning to update parameters of an MPC algorithm based on feedback from applying control inputs. Further, several works develop and compare various deep reinforcement learning methods to find and track optimal trajectories \cite{Remonda2021, deBruin2018, weiss2020}. 

In the context of multi-agent racing works, both optimization and learning-based control approaches are also used. Li et. al \cite{Li2021} develop a mixed-integer quadratic programming formulation for head-to-head racing with realistic collision avoidance but concede that this formulation struggles to run in real-time. Another study proposes a real-time control mechanism for a game with a pair of racing drones \cite{spica2020real}. This work provides an iterative-best response method while solving an MPC problem that approximates a local Nash equilibrium. It is eventually extended to automobile racing \cite{Wang2019} and multi-agent scenarios with more than two racers, but they do not consider teams \cite{Wang2021}. A fast, real-time MPC algorithm to make safe overtakes is presented in \cite{He2021}, but the method does not consider adversarial behavior from the opposing players. Similar to the single-agent racing case, some studies use deep learning methods to train neural network based controllers \cite{Schwarting2021, Song2021}. Again, all of these studies do not consider racing rules except for collision avoidance without the nuances of responsibility nor do they incorporate team-based objectives.

However, Wurman et al. \cite{sonyai} develop an autonomous racing controller using deep reinforcement learning that considers the rules of racing beyond collision avoidance. Their controller outperforms expert humans while also adhering to proper racing etiquette. It is the first study to consider nuanced safety and fairness rules of racing and does so by developing a reward structure that trains a controller to understand when it is responsible for avoiding collisions, and when it can be more aggressive. They do not encode the rules directly in their model. Instead, they refer to human experts to evaluate the behavior of their trained deep learning controllers to adjust parameters that affect the aggressiveness of their controller. Their control design is fully learning-based and does not involve explicit path planning or hierarchical reasoning. In addition, although this paper models more realistic racing behavior in multi-agent racing, it also still lacks consideration of cooperative objectives amongst teams of racers.

Hierarchical game-theoretic reasoning is a method that has been previously studied in the context of autonomous driving. A hierarchical racing controller is introduced in \cite{LinigerThesis} by constructing a high-level planner with simplified dynamics to sample sequences of constant curvature arcs and a low-level planner to use MPC to track the arc that provided the furthest progress along the track. A two-level planning system is developed in \cite{Fisac2019} to control an autonomous vehicle in an environment with aggressive human drivers. The upper-level system produces a plan to be safe against the uncertainty of the human drivers in the system by using simplified dynamics. The lower-level planner implements the strategy determined by the upper level-planner using precise dynamics. Similarly, Moghadam and Elkaim \cite{Moghadam2019} also study hierarchical reasoning decision making in highway driving. They construct a high-level planner using a trained reinforcement-learning policy to determine lane changing plans to safely pass other drivers. The lane changing plans are shared with low-level controllers to execute those actions. These papers have established the power of hierarchical reasoning in autonomous driving, but they have only applied it in a non-adversarial context. However, in the autonomous racing scenario, other participants in the system have competing objectives, which complicates how the hierarchical abstraction must be constructed.

Cooperative control in multi-agent systems is also an area that has been extensively studied and applied to many domains. For example, in a review by Wang et al. \cite{wang2017cooperative}, the authors compile several papers that apply multi-agent cooperative control to some theoretical problems such as path covering, target tracking, and distributed consensus. We have also seen it applied at an application specific scope such as coordinated traffic control \cite{Abdoos2021} and robot soccer \cite{hwang2004cooperative}. However, as far as our research shows, cooperative control for autonomous racing has not been explored previously.

While research in autonomous racing has much more literature across all of the components of development\cite{litreview}, almost all of the works lack joint consideration of two important components that would allow it to more closely resemble real-life racing: rules in addition to basic collision avoidance and teams of players. This project aims to fill that gap and show how game theoretic hierarchical reasoning is a powerful tool for designing controllers in multi-agent systems involving complex rules and objectives.

\section{Team-based Multi-Agent Racing Game Formulation}
\begin{figure*} [!ht]
  \centering
  \includegraphics[width=\textwidth]{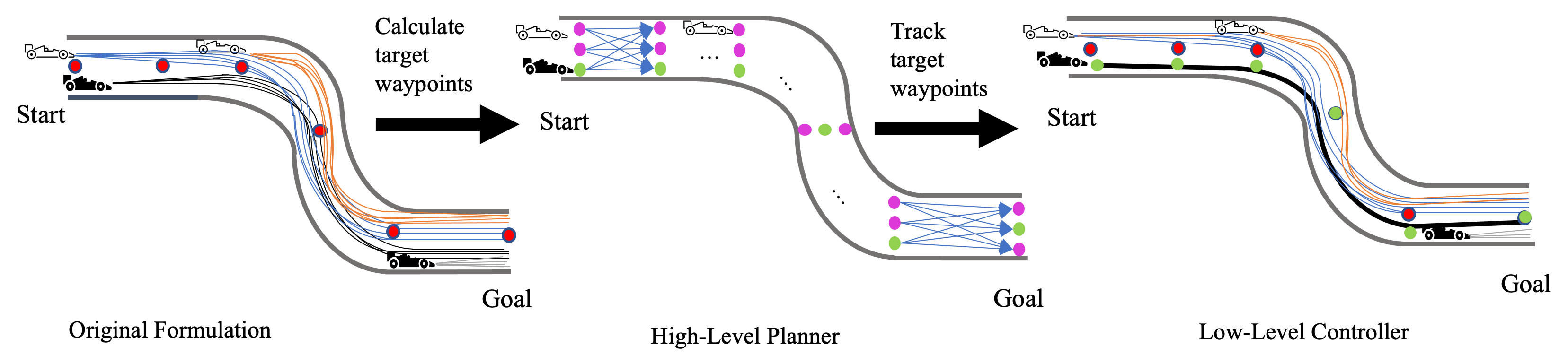}
  \caption{We show an overall view of our planning algorithm with the perspective of the black car at the start. There are many seemingly reasonable trajectories in the general game (left). The high-level planner constructs a discretized approximation, which only considers nearby players (middle). The low-level controller tracks the sequence of target waypoints calculated by the high-level planner in green, which is represented by a continuous trajectory in black (right).}
  \label{fig:overall_control}
\end{figure*}
To motivate the proposed control design, we first outline a dynamic game formulation of a multi-agent racing game involving teams. Table \ref{tab:symbols}, provided in the Appendix, lists all of the variables and functions referenced in this formulation and the formulations introduced in the following sections. 

Let there be a set $N$ of players racing over $T$ discrete time steps in $\mathcal{T} = \{1, ..., T\}$. We introduce a set $M$ consisting of mutually exclusive subsets of players in $N$. Each of the sets in $M$ represents a team of players whose objectives involve an incentive to collectively finish ahead of the players in the other teams in the race. We define a racetrack using a sequence of $\tau$ checkpoints along its center line, $\{c_i\}_{i=1}^{\tau}$, whose indices are in a set $C=\{1,..., \tau\}$. Each player's continuous state (including, e.g., position, speed, or tire wear) is denoted as $x^i_t \in X \subseteq \mathbb{R}^n$, and control is denoted as $u^i_t \in U \subseteq \mathbb{R}^k$. We also introduce a pair of discrete state variables $r^i_t \in C$ and $\gamma^i \in \mathcal{T}$. The index of the latest checkpoint passed by player $i$ at time $t$ is $r^i_t$, and it is computed by function $p: X\rightarrow C$. The earliest time when player $i$ reaches the final checkpoint is $\gamma^i$, i.e. $r^i_{\gamma^i} = \tau$. We define a multiplier $\zeta \in [0,1]$ to balance a player's emphasis on its team's performance vs. its own performance. Using these definitions, the objective for each Player $i$ on a team $\mu$ is expressed as:
\begin{equation} \label{eq:gen_obj}
    \min_{u^i_0, ..., u^i_T} \gamma^i+\zeta (\sum_{j \in \mu \setminus i} \gamma^j) - \frac{(1+\zeta(|\mu|-1))\sum_{j \in N \setminus \mu }\gamma^j}{|N|-|\mu|} 
\end{equation}
In effect, the players aim to minimize their own time and the sum of the times of their teammates to reach the final checkpoint while maximizing the sum of the times it takes all other players to reach the final checkpoint. While this is not the precise definition of winning the race by coming in first place, it is a practical and smooth approximation to that idea. Also, note that if we assume players act independently, i.e. setting $\zeta = 0$ or $|\mu|=1 \,\, \forall \, \mu \in M$, the objective is equivalent to the N-player generalization to 2-player formulation presented in prior work \cite{thakkarprior}. 

Player $i$'s state $x^i_t$ and control $u^i_t$ are governed by known dynamics $f^i$. The core dynamics of the game, including those managing the previously introduced discrete state variables, for all players $j \in N$ are as follows:
\begin{equation} \label{eq:gen_dyn}
    x^j_{t+1} = f^i(x^j_t, u^j_t), \quad \forall \;\; t \in \mathcal{T}
\end{equation}
\begin{equation} \label{eq:gen_idx_map}
    r^j_{t+1} = p(x^j_{t+1}, r^j_t), \quad \forall \;\; t \in \mathcal{T}
\end{equation}
\begin{equation} \label{eq:gen_init_idx}
    r^j_{1} = 1
\end{equation}
\begin{equation} \label{eq:gen_reach_goal}
    r^j_{T} = \tau
\end{equation}
\begin{equation} \label{eq:gen_goal_time}
    \gamma^j = \min \{t \, | \, r^i_t = \tau \wedge t \in \mathcal{T} \}
\end{equation}

In addition to the individual dynamics, we introduce constraints modeling the rules of the game. To ensure that the players stay within the bounds of the track we introduce a function, $q: X \rightarrow \mathbb{R}$, which computes a player's distance to the closest point on the center line. This distance must be limited to the width of the track $w$. Therefore, for all $t \in \mathcal{T}$ and $j \in N$:
\begin{equation} \label{eq:gen_idx_dist}
    q(x^j_{t}) \leq w
\end{equation}

Next, we define the collision avoidance rules. We evaluate if player $i$ is ``behind" player $j$, and depending on the condition, the distance between every pair of players, computed by the function $d: X \rightarrow \mathbb{R}$, is required to be at least $s_1$ if player $i$ is behind another player $j$ or $s_0$ otherwise. For all $t \in \mathcal{T}$, $j \in N$, and $k \in N \setminus \{j\}$ these rules are expressed by the constraint:
\begin{equation} \label{eq:gen_coll_avoid}
    d(x^i_{t}, x^j_t) \geq  \begin{cases} s_1 & \text{player} \, i \, \text{behind player}\,j\\
    s_0 & \text{otherwise}  \end{cases}
\end{equation}

Finally, players are limited in how often they may change lanes depending on the classification of part of the track they are located at. We assume that there are $\lambda \in \mathbb{Z^+}$ lanes across all parts of the track. If the player's location on the track is classified as a curve, there is no limit on lane changing. However, if the player is at a location classified as a straight, it may not change lanes more than $L$ times for the contiguous section of the track classified as a straight. We define a set $\mathcal{S}$ that contains all possible states where a player is located at a straight section. We also introduce a function $z: X \rightarrow \{1, 2, ..., \lambda\}$ that returns the lane ID of a player's position on the track. Using these definitions, we introduce a variable $l^j_t$ calculated by the following constraint for all $t \in \mathcal{T}$ and $j \in N$:
\begin{equation} \label{eq:gen_lane_var}
    l^j_{t} =  \begin{cases} l^j_{t-1} + 1 & \mathds{1}_{x^j_t \, \in \mathcal{S}} = \mathds{1}_{x^j_{t-1} \in \mathcal{S}}\wedge z(x^j_t) \neq z(x^j_{t-1}) \\
    0 & \text{otherwise}  \end{cases}
\end{equation}
$l^j_{t}$ represents a player's count of ``recent'' lane changes over a sequence of states located across a contiguous straight or curved section of the track. However, the variable is only required to be constrained if the player is on a straight section of the track. Therefore, the following constraint must hold for all $t \in \mathcal{T}$ and $j \in N$ and if $x^j_t \, \in \mathcal{S}$:
\begin{equation} \label{eq:gen_lane_lim}
    l^j_{t} \leq  L
\end{equation}

Most prior multi-agent racing formulations do not include the complexities introduced through constraints \eqref{eq:gen_coll_avoid}-\eqref{eq:gen_lane_lim}, \cite{Wang2019, Wang2021, He2021, Schwarting2021, Song2021}. Instead, they usually have a similar form regarding continuous dynamics and discrete checkpoints \eqref{eq:gen_dyn}-\eqref{eq:gen_goal_time}, and their rules only involve staying on track \eqref{eq:gen_idx_dist} and collision avoidance with a fixed distance for all players regardless of their relative position. However, in real-life racing, there exist complexities both in the form of mutually understood unwritten rules and explicit safety rules \cite{racingrules}. As a result, we account for two of the key rules that ensure the game remains fair and safe:
\begin{enumerate}
    \item There is a greater emphasis on and responsibility of collision avoidance for a vehicle that is following another \eqref{eq:gen_coll_avoid}.
    \item The player may only switch lanes $L$ times while on a straight section of the track \eqref{eq:gen_lane_var}-\eqref{eq:gen_lane_lim}.
\end{enumerate}

The first rule ensures that a leading player can make a decision without needing to consider an aggressive move that risks a rear-end collision or side collision while turning from the players that are following. This second rule ensures that the leading player may not engage in aggressive swerving across the track that would make it impossible for a player that is following the leader to safely challenge for an overtake. While there exist functions to evaluate these spatially and temporally dependent constraints, their discrete nature makes them difficult to differentiate. As a result, most state-of-the-art optimization algorithms may not be applicable or may struggle to find a solution in real time.
\section{Hierarchical Control Design}
In the example given in the introduction, there are three main strategies that seem reasonable. However, there is also an infinite set of strategies that lie between these three options. Because of this, it is computationally infeasible to consider and compare all of the possible strategies in terms of their satisfiability with respect to the rules in Equations \eqref{eq:gen_dyn} - \eqref{eq:gen_lane_lim} and their optimality with respect to the objective in Equation \eqref{eq:gen_obj}. The rules and the objective involve nonlinear functions over both continuous and discrete variables, which makes it unlikely that a mixed-integer nonlinear programming algorithm could be used to solve the game at high frequencies for precise control. This inherent challenge motivates the use of methods such as deep reinforcement learning or short receding horizons. However, we do not solely rely on these methods as seen in previous works.

We propose a two-level hierarchical control design involving two parts that work to ensure the rules are followed while approximating long-term optimal choices. The high-level planner transforms the general formulation into a game with discrete states and actions where all of the discrete rules are naturally encoded. The solution provided by the high-level planner is a series of discrete states (i.e waypoints) for each player, which satisfies all of the rules. Then, the low-level planner solves a simplified version of the racing game. The simplified version has an objective that places greater emphasis on tracking a series of waypoints and smaller emphasis on the original game-theoretic objective and a reduced version of the rules. Therefore, this simplified game can be solved by an optimization method in real-time or be trained in a neural network when using a learning-based method. 

This control design assumes that if the series of waypoints produced by the high-level planner is guaranteed to follow the rules, then the control inputs generated by the waypoint tracking low-level planner will also satisfy the rules of the original game when applied to the actual underlying system. Figure \ref{fig:overall_control} visualizes how overall control architecture is applied.
\subsection{High-Level Tactical Planner}
The high-level planner constructs a turn-based discrete, dynamic game that is an approximation of the general game \eqref{eq:gen_obj}-\eqref{eq:gen_lane_lim}. In the following subsections, we discuss how we discretize the state space, simplify the dynamics, and solve the game. 

\subsubsection{State Space Discretization}
We begin by constructing the discrete abstraction of the state space from the original formulation. We do not explicitly specify any components of players' states when defining the original formulation because it is agnostic to the vehicle dynamics model being considered. However, including variables computed by constraints \eqref{eq:gen_idx_map} and \eqref{eq:gen_lane_var}, we assume each player's state in the original formulation at least consists of following five variables as they are the only ones modeled in our dynamics and state representation: position, velocity, number of ``recent" lane changes, tire wear, last passed checkpoint index.
    
We specify the play order so that the discrete game progresses by players making choices at the checkpoints indexed by elements of $C$ rather than at each time-step from $\mathcal{T}$. This transformation is natural to consider because all players must ultimately pass all of the checkpoints in order. As a result, the turns of the discrete game and players' states in the discrete game are indexed by their last passed checkpoint, and the time step becomes a variable in the discrete game state. Furthermore, indexing by the checkpoints also produces a natural discretiziation for the position state variable in the original formulation. Around each checkpoint, we select $\lambda$ (which is the number of lanes) discrete locations along the line perpendicular to the direction of travel. Each of the $\lambda$ locations evaluates to a unique lane ID on the track when passed into function $z(\cdot)$ defined in the general formulation. Therefore, we represent a player's position in discrete game by its lane ID for a given index of the game state i.e., the last passed checkpoint. This choice enables us to naturally encode the rules governing players' lanes and ensures that every location considered in the discrete game remains within the bounds of the track.

\begin{figure}
  \centering
  \includegraphics[width=0.48\textwidth]{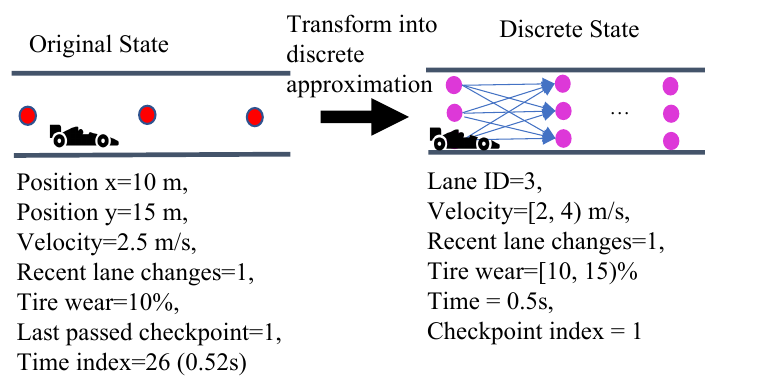}
  \caption{An example of a player's state in the original game (top) is converted into our discrete game approximation (bottom). The position is converted into a lane ID and checkpoint index. Velocity and tire wear are projected into ranges of some fixed size. The time step is reduced to lower, finite precision time state in the discrete game. The recent lane changes state variable remains unchanged because it is inherently discrete.}
  \label{fig:state_discrete}
\end{figure}

The remaining components of players' states are either already discrete valued (such as the count of ``recent lane changes"),  represented in the form of discrete ranges, or rounded to a finite precision. For example, instead of considering real number value for a Player $i$'s velocity from its state $x^i_v=\SI{2.5}{\meter\per\second}$ in the original game, the discrete representation would simply be $v^i \in [2, 4) \si{\meter\per\second}$ meaning that the continuous velocity falls within the given range. TThese ranges are predetermined based on the size of the state space that is manageable for the computational resources. The overall components of Player $i$'s discrete state consist of lane ID $a^i_k$, velocity range $v^i_k$, number of ``recent" lane changes $l^i_k$, tire wear proportion $e^i_k$, and time $t^i_k$ where $k$ is the index of the state and the last passed checkpoint associated with the state. Figure \ref{fig:state_discrete} shows how the continuous space of the track with checkpoints (in red) is transformed into discrete locations associated with a unique lane ID at each checkpoint (in purple). It also illustrates how the state in the original game (left) is transformed into the discrete game representation (right).

\subsubsection{Dynamics Abstraction}
Given the state space transformation, we discuss the dynamics of the discrete game. The players' actions are defined by pairs of lane ID and target velocity range for the upcoming checkpoint. Therefore, we approximate the original dynamics using one-dimensional equations of motion to determine the time it would take to transition from one checkpoint to the next. Using those calculations, we also estimate the remaining state variables or rule out the actions if they are dynamically infeasible in our approximation. 

To calculate updates for the elapsed time state $\delta t_k$, we first use the known track parameters (such as turning radius or lane width) to estimate the travel distance $d$ between a player's lane at the current checkpoint $c_k$ to the target lane in the subsequent checkpoint $c_{k+1}$. If the track between two checkpoints is a straight, the Euclidian is used to estimate the distance to travel based on the lane width $w_l$, difference between the player's initial lane and target lane, and the straight line distance between the location of the checkpoints $\upsilon_{k, k+1}$. If the the track between the two checkpoints is a curve, then we calculate a coarse estimate of the distance by averaging the radius of the turn for the player's lane at the initial checkpoint $r_k$ and the radius of the turn for the player's target lane at the next checkpoint $r_{k+1}$ and multiply it by the central angle of the turn $\theta_k$. These calculations are summarized below:
\begin{equation} \label{eq:dist_calc}
    d = \begin{cases}
    \sqrt{(w_l|a_k - a_{k+1}|)^2 + \upsilon_{k, k+1}^2} &  \text{if } k \in \mathcal{S} \\
    \frac{r_k + r_{k+1}}{2}\theta_k & \text{otherwise}
    \end{cases}
\end{equation}

Once the distance $d$ is known, we use the average of the velocity range at the initial checkpoint $\bar{v}_k$, average of the velocity range at the target checkpoint $\bar{v}_{k+1}$, and known parameters of the vehicle to estimate minimum time it takes to travel a given distance. The known parameters of the vehicle include maximum acceleration $a$, maximum deceleration $b$, maximum velocity $v_{\text{max}}$, and a derived quantity denoted as maximum allowed velocity $v_*$. We derive $v_*$ to be the minimum of $v_{\text{max}}$ and a calculation derived from lateral acceleration sustained when driving in a circle. This calculation relies on the tire wear proportion at the initial checkpoint $e_k$, track radius at the initial checkpoint $r_k$, and additional known vehicle parameters of maximum allowed lateral acceleration $a_{\text{max}}$ and minimally feasible lateral acceleration $a_{\text{min}}$ (i.e. lateral acceleration the vehicle can sustain regardless of tire wear state). The equation to compute $v_*$ is the following:
\begin{equation}
    v_* = \min \{\sqrt{(a_{\text{max}} - (a_{\text{max}}-a_{\text{min}})e_k)r_k}, v_{\text{max}} \}
\end{equation}

Once $v^*$ is known, we have all of the components to estimate the minimum time to travel a distance using equations of one dimensional motion. In the calculation, we enforce that $\bar{v}_{k+1} \leq v_*$ and disregard all actions that violate this constraint because they would not obey the lateral acceleration or top speed limitations of the vehicle in our approximation. In addition, we verify it is possible to accelerate or decelerate from $\bar{v}_k$ to $\bar{v}_{k+1}$ within the distance $d$. If that is not possible, then the action with average target velocity $\bar{v}_{k+1}$ is also disregarded. For the remaining cases, we use the following calculation to determine the time update $\delta t_k$ for an agent going from initial velocity $\bar{v}_k$ to target velocity $\bar{v}_{k+1}$, maximum acceleration $a$, maximum braking $b$, and maximum allowed velocity $v_*$:

\begin{equation} \label{eq:time_update}
    \delta t_{k} = \begin{cases}
    \frac{v_*-\bar{v}_k}{a} + \frac{v_*-\bar{v}_{k+1}}{b} + \frac{d-\frac{v_*^2-\bar{v}_k^2}{2a} - \frac{v_*^2-\bar{v}_{k+1}^2}{2b}}{v_*} 
    & \\ \qquad\qquad \text{if } v_* \geq \bar{v}_k \; \wedge 
     \quad \frac{d-\frac{v_*^2-\bar{v}_k^2}{2a} - \frac{v_*^2-\bar{v}_{k+1}^2}{2b}}{v_*} \geq 0 \\ & \\
    \frac{\bar{v}_k -v_*}{b} + \frac{v_*-\bar{v}_{k+1}}{b} + \frac{d-\frac{\bar{v}_k^2 -v_*^2}{2b} - \frac{v_*^2-\bar{v}_{k+1}^2}{2b}}{v_*} 
    & \\ \qquad\qquad  \text{if } v_* < \bar{v}_k \;\wedge 
    \quad \frac{d-\frac{\bar{v}_k^2 -v_*^2}{2b} - \frac{v_*^2-\bar{v}_{k+1}^2}{2b}}{v_*} \geq 0 \\ & \\
    \frac{\sqrt{\frac{-2dba -b\bar{v}_k^2 -a\bar{v}_{k+1}^2}{-a-b}}-\bar{v}_{k}}{a} + \frac{\sqrt{\frac{-2dba -b\bar{v}_k^2 -a\bar{v}_{k+1}^2}{-a-b}}-\bar{v}_{k+1}}{b}
    & \\ \qquad\qquad \text{if } v_* \geq \bar{v}_k \; \wedge 
     \quad \frac{d-\frac{v_*^2-\bar{v}_k^2}{2a} - \frac{v_*^2-\bar{v}_{k+1}^2}{2b}}{v_*} < 0 \\ & \\
    \text{action ruled out} \qquad\qquad\qquad\qquad\;\;\;\; \text{otherwise}
    \end{cases}
\end{equation}

This calculation assumes that the player accelerates or brakes to reach $v_*$ from $\bar{v}_k$, maintains that speed for as long as possible until the player must brake to hit $\bar{v}_{k+1}$ if $\bar{v}_{k+1} \neq v_*$. If there is not enough distance to perform this maneuver and $\bar{v}_k \leq v_*$, we calculate the highest velocity the player can reach given we must end at the target velocity $\bar{v}_{k+1}$ within the distance $d$. All other possible maneuvers would violate the approximated dynamical limitations of the vehicle and are ruled out of the set of allowed actions  player may choose. We also use the time state update \eqref{eq:time_update} to estimate collision avoidance. If a player chooses a lane that a prior player has already selected for its turn and the difference in the time states for these players would be smaller than some time-window if the action is applied, then the action is disregarded for being a high risk of collision.

Finally, in order to calculate the tire wear state update, we use different calculations for the straight or curve sections of the track. If the track between the checkpoints is a straight, we multiply a tire wear factor parameter $L_{\text{straight}}$ associated with driving straight with the distance of the straight $d$. When the track between the checkpoints is a curve, we multiply the tire wear factor parameter $L_{\text{curve}}$ associated with driving on a curve, the distance of the curve $d$, and an estimate for the average lateral acceleration achieved by hitting the target velocity $\bar{v}_{k+1}$ calculated using equations of circular motion. The tire wear update $\delta e_k$ is calculated as follows:
\begin{equation}
\delta e_k = \begin{cases}
    dL_{\text{straight}} & \text{if } k \in \mathcal{S} \\
    \frac{2dL_{\text{curve}}\bar{v}_{k+1}^2}{r_k + r_{k+1}} & \text{otherwise}
\end{cases}
\end{equation}

For both the time and tire wear states, the updates are added to the initial state and projected back into their discrete ranges or rounded to the finite precision. Note that all of the known parameters used in our calculations are standard in most vehicle dynamics models except for tire wear related parameters\cite{Rajamani2011vehicle}. We emphasize this note because our high-level planner is designed to be agnostic to the underlying dynamics model. If tire wear is not modeled, one can just assume that $e_k$ is always zero, and the remaining calculations are left unchanged or unused without impacting the discrete game implementation.

As briefly mentioned earlier, this action space abstraction also allows us to easily evaluate or prevent actions where rules of the game would be broken. By limiting positional choices to fixed locations across checkpoints, we ensure that the players always remain on track \eqref{eq:gen_idx_dist}. Moreover, the players' actions are dismissed if they would violate the limit on the number of lane changes by simply checking whether choosing a lane would exceed their limits or checking if the location is a curve or straight \eqref{eq:gen_lane_lim}. Finally, other actions that could cause collisions are also dismissed by assuming that if two players reach the same lane at a checkpoint and have a small difference (e.g. \SI{0.1}{\second}) in their time states, there would be a high risk of collision \eqref{eq:gen_coll_avoid}.

The game is played with each player starting at the initial checkpoint, and it progresses by resolving all players' choices and state updates one checkpoint at a time. The order in which players take their turns is determined by the player with the smallest time state at each checkpoint. This means that players who arrived at a checkpoint earlier get to make their choices before players who arrived later, and that players who arrive later get to observe the actions of the players who arrived earlier. It is also possible to use a time-step ordering in this model, i.e. the order determined by the one having the smallest time state across all checkpoints, which produces a more precise representation of the flow of information. However, we discuss in the following subsection that the players construct the game only considering opponents within a small radius. As a result, both methods yield similar, if not identical, order of player turns and checkpoints. 

\subsubsection{Game Solution}
The objective of the discrete game is to minimize the difference between one's own time state aggregated with the sum of one's teammates time states at the final checkpoint and the sum of the time states of all other players at the final checkpoint just like the original formulation \eqref{eq:gen_obj}. This objective is to minimize the following equation, which is returned as the score of the game for player $i$ on team $\mu$ at terminating state of the game once all players have reached the final checkpoint:
\begin{equation}
 t^i_\tau + \zeta(\sum_{k \in \mu \setminus i} t^k_\tau) - \frac{(1+\zeta(|\mu|-1))\sum_{j \in N \setminus \mu} t^j_\tau}{|N|-|\mu|}
\end{equation}
Although the discrete game model is simpler than the original formulation, the state space grows exponentially as the number of players, actions, and checkpoints increases. Therefore, selecting the density of the checkpoints is important parameter in the model. If we introduce too many checkpoints to better model the track, the dynamics abstractions might become too great of a simplification to be useful and the state space would just be unmanageable. On the other hand, too sparse of a checkpoint setup limits the possibilities of meaningful strategic plans. Our model sets the checkpoints to be \SI{10}{\meter}-\SI{15}{\meter} apart. 

We solve the game in a receding horizon manner by assuming the ``final'' checkpoint is 8 checkpoints ahead of the ego player's current checkpoint and only considering opponents within a nearby radius. It is possible that nearby players may not be at the same checkpoint as the ego player, even if they are within the nearby radius. To set up the initial discrete approximations of the opponents, we can use the complete state information to determine the components of the discrete approximation, except for the time state component. To determine the time state component of nearby opponent players, we additionally assume that all players have knowledge of when every other player has passed each checkpoint. Using this knowledge, we compute the time difference at the last checkpoint that both the ego player and the nearby opponent being constructed have passed, and set that difference as the initial time state of the opponent. The remaining steps in the initialization and updates of the opponent's state are the same as those discussed in previous sections.

Our choice of horizon at 8 checkpoints allows us to plan further into the future than an MPC-based continuous state/action space controller can handle in real time. For example, the distance covered by 8 checkpoints in our horizon is upwards of 80 meters while the MPC-based continuous controller only plans up to 25-30 meters ahead in \cite{Wang2019, Wang2021}. We use the Monte Carlo tree search (MCTS) algorithm \cite{mcts} to produce an approximate solution for our game in real time. The solution from applying MCTS is a series of waypoints in the form of target lane IDs (which can be mapped back to positions on track) and the target velocities at each of the checkpoints for the ego player and estimates of the best response lanes and velocities for the other players.

Our discrete game abstraction is an useful representation of real-life racing because it captures the natural discretization that exists in the rules and strategy. For example, there are rules governing the lane-changing, which also involve conditions on how different parts of the track are classified. Intuition also suggests that frequent changes in direction are both suboptimal and unsafe, because they destabilize vehicles and increase the risk of collision. As a result, drivers do not frequently change directions without good reason. Instead, they make strategic choices about which part of the track to occupy at discrete locations, represented by the checkpoints in the discrete abstraction. The rest of their effort is focused on reaching these planned locations. Our hierarchical model is based on these ideas. The high-level tactical planner produces a series of target waypoints at each checkpoint that satisfy the rules, and the low-level path planner determines control inputs to reach those waypoints.

\subsection{Low-Level Controller}
The low-level controller is responsible for producing the control inputs, so it must operate in real-time. Because we have a long-term plan provided by the high-level planner that has considered some of the complex rules, we can formulate a reduced version of the original game for our low-level planner. The low-level game is played over a shorter horizon compared to the original game of just $\delta$ discrete time steps in $\hat{\mathcal{T}} = \{1, ..., \delta\}$. We assume that the low-level planner for player $i$ has received $k$ waypoints, $\psi^i_{r^i_{1}}, ..., \, \psi^i_{r^i_{1} + k}$, from the high-level planner, and player $i$'s last passed checkpoint $r^i_*$. 

The low-level objective involves two components. The first component is to maximize the difference between the sum of its own progress and its team's progress and sum other agents' progress at the end of $\delta$ steps where progress is indicated by the last passed checkpoint index variable $r^i_{\delta}$. The second component is to minimize its tracking error, $\eta^i_y$, of every passed waypoint $\psi^i_{r^i_{1}+y}$. The former component influences the player to pass as many checkpoints as possible and aid its teammates in passing their checkpoints, which overall, suggests helping its team reach the final checkpoint, $c_\tau$, as quickly as possible. The latter influences the player to be close to the calculated high-level waypoints when passing each of the checkpoints. The objective also includes a multiplier $\alpha$ that balances the emphasis of the two parts of the objective. The objective for Player $i$ is written as follows:

\begin{multline} \label{eq:ll_obj}
    \min_{u^i_{1}, ..., u^i_{\delta}} \Big(\frac{((1+\zeta(|m|-1))\sum^N_{j \in N\setminus \mu}r^j_{\delta}}{|N|-|m|} -  r^i_{\delta} - \zeta\sum_{j \in \mu \setminus i}r^j_\delta\Big) \\ + 
    \alpha \sum_{c={r^i_{1}}}^{{r^i_{1}}+k} \eta^i_c
\end{multline}

The players' continuous state dynamics, calculations for each checkpoint, and constraints on staying within track bounds \eqref{eq:ll_dyn}-\eqref{eq:ll_pos_dist} are effectively the same as the original formulation. For all players $j \in N$, the following must hold:
\begin{equation} \label{eq:ll_dyn}
    x^j_{t+1} = f(x^j_{t}, u^j_t), \quad \forall \;\; t \in \hat{\mathcal{T}}
\end{equation}
\begin{equation} \label{eq::ll_pos}
    r^j_{t+1} = p(x^j_{t+1}, r^j_t), \quad \forall \;\; t \in \hat{\mathcal{T}}
\end{equation}
\begin{equation} \label{eq::ll_pos_init}
    r^j_{1} = r^j_*
\end{equation}
\begin{equation} \label{eq:ll_pos_dist}
    q(x^m_{t}) \leq w, \quad \forall \;\; t \in \hat{\mathcal{T}}
\end{equation}

The collision avoidance rules are simplified to just maintaining a minimum separation $s_0$ as the high-level planner would have already considered the nuances of rear-end collision avoidance responsibilities outlined in \eqref{eq:gen_coll_avoid}. As a result, we require the following constraint to hold for all $t \in \hat{\mathcal{T}}$, $j \in N$, and $k \in N \setminus \{j\}$:
\begin{equation} \label{eq:ll_coll_avoid}
    d(x^j_{t}, x^k_t) \geq s_0
\end{equation}

Finally, we define the dynamics of the waypoint error, $\eta^i_y$, introduced in the objective. It is equivalent to the accumulated tracking error of each target waypoint that player $i$ has passed using a function $h: X\times X \rightarrow \mathbb{R}$ that measures the distance. If a player has not passed a waypoint, then the error variable indexed by that waypoint is set to 0. Its dynamics are expressed by the following constraint:

\begin{multline} \label{eq:ll_wp_err}
    \eta^i_y = \begin{cases} \sum_{t}^{\Hat{\mathcal{T}}} h(x^i_t, \psi^i_{c})  & \text{if } \exists \; r^i_t \geq y \\
    0 & \text{otherwise}
    \end{cases} \\ \forall \; y \in \{r^i_{1}, ..., r^i_{1} + k\}
\end{multline}

This simplified formulation in Equations \eqref{eq:ll_obj}-\eqref{eq:ll_wp_err} is similar to the general formulation in Equations \eqref{eq:gen_obj}-\eqref{eq:gen_lane_lim}. The objective, in addition to involving the notion of maximizing the progress of one's team, influences the controller to stay as close to and pass as many checkpoints as possible (indicated by the term multiplied by $\alpha$). Furthermore, the constraints introduced by the complex fairness and safety rules in the original formulation are simplified or dropped because we assume them to be considered by the high-level planner. 

However, it is important to note that the high-level and low-level planners in our system run independently and concurrently, which can sometimes lead to violations of the rules of the original game. This is particularly likely to happen when the high-level planner constructs a plan based on outdated state information. For example, the low-level planner may cause a player to take an unexpected evasive action that exceeds its lane changing limit while the high-level planner is still calculating the next sequence of waypoints before this action was taken. Because the low-level formulation does not explicitly account for this rule, the player may attempt to change lanes again, resulting in a rule violation. Despite this limitation, our empirical results show that the hierarchical planners are more consistent in following the rules compared to other methods, as they generally stay close to their original high-level plan.

We consider two separate computational methods to solve this low-level formulation. The first method develops a reward and an observation structure to represent this simplified formulation for a multi-agent reinforcement learning (MARL) algorithm to train a policy that serves as a controller. The second method further simplifies the low-level formulation into a linear-quadratic Nash game (LQNG) to compute short-horizon control inputs. We consider two low-level methods to study the versatility of our high-level tactical planner. Because the low-level planners still incorporate game-theoretic planning, it enables us to compare our hierarchical architecture with our high-level planner against control architectures that use a pre-computed optimal racing line with local game-theoretic reasoning. 

\subsubsection{Multi-Agent Reinforcement Learning Controller}
Designing the MARL controller primarily involves shaping a reward structure that models the low-level formulation. While we provide a high-level description of the reward and penalty behaviors below, the \hyperref[sec:app]{Appendix} includes specific details about the reward functions and when they are applied. The \hyperref[sec:app]{Appendix} also contains details regarding the neural network architecture and details about the training procedure. In general, the RL agent is rewarded for the following behaviors that would improve the objective function from the low-level formulation \eqref{eq:ll_obj}:
\begin{itemize}
    \item Passing a checkpoint with an additional reward for being closer to the target lane and velocity.
    \item Minimizing the time between passing two checkpoints.
    \item Passing as many checkpoints in the limited time.
    \item Helping one's teammates pass as many checkpoints in the limited time with higher rewards if one's teammates pass the checkpoint ahead of opposing teams' players.
\end{itemize}
On the other hand, the agent is penalized for actions that would violate the constraints:
\begin{itemize}
    \item Swerving too frequently on straights \eqref{eq:gen_lane_lim}.
    \item Going off track or hitting a wall \eqref{eq:ll_pos_dist}.
    \item Colliding with other players \eqref{eq:ll_coll_avoid} with additional penalty if the agent is responsible for avoidance \eqref{eq:gen_coll_avoid}. 
\end{itemize}

The rewards capture our low-level formulation objective \eqref{eq:ll_obj} to pass as many checkpoints as possible while closely hitting the lane and velocity targets \eqref{eq:ll_wp_err}. The penalties capture the on-track \eqref{eq:ll_pos_dist} and collision avoidance \eqref{eq:ll_coll_avoid} constraints. However, the penalties also reintroduce the original safety and fairness from the original general game that were simplified away from the low-level formulation \eqref{eq:gen_coll_avoid} and \eqref{eq:gen_lane_lim}. Because these rules are inherently met by satisfying the objective of reaching the high-level planner's waypoints, their penalties have the weights set much lower than other components of the reward/penalty structure. Nevertheless, we still incorporate the original form of these penalties to reinforce against the possibility that the ego player might be forced to deviate far away from the high-level plan.

The agents' observations include perfect state information (velocity, relative position, tire wear, lane change counts, and last passed checkpoint) of all players and local observations consisting of 9 LIDAR rays spaced over a 180\textdegree{} field of view centered in the direction that the player is facing with a range of up to \SI{20}{\meter}. Lastly, the agents also observe the relative location of the $k$ upcoming target waypoints, $\psi^i_{r^i_{1}}, ..., \, \psi^i_{r^i_{1} + k}$ as it is rewarded for reaching those waypoints.

\subsubsection{Linear-Quadratic Nash Game Controller}
Our second low-level approach solves an LQNG using the coupled Riccati equations \cite{basar}. This method involves further simplifying the low-level formulation into a structure with a quadratic objective and linear dynamics. The continuous state is simplified to just four variables: $x$ position, $y$ position, $v$ velocity, and $\theta$ heading. The control inputs $u^i_t$ are also explicitly broken into acceleration, $a^i_t$, and yaw-rate, $e^i_t$. The planning horizon is reduced to $\Bar{\delta}$ where $\Bar{\delta} \ll \delta < T$. To construct our quadratic objective for player $i$, we break it into three components. The first is to minimize the squared distance to the upcoming target waypoint from the high-level planner $\Bar{\psi}^i$ calculated by the following function of some weight parameters $\rho_1, \rho_2, \text{and } \rho_3$:
\begin{multline} \label{eq:lqng_obj1}
\upsilon^i(\Bar{\psi}^i, \rho_1, \rho_2,\rho_3) =  \sum_{t = 1}^{\Bar{\delta}} (\rho_1((x^i_{t} - \Bar{\psi}^i_x)^2 + (y^i_{t} - \Bar{\psi}^i_y)^2) \\   \quad + \rho_2 (v^i_{t} - \Bar{\psi}^i_v)^2 
 + \rho_3 (\theta^i_{t} - \Bar{\psi}^i_\theta)^2)
\end{multline}

The second component is to maximize or minimize each of the other player's distances from the location of their estimated target waypoints $\Bar{\psi^j}$. If the other player is on the ego's team, then ego would like to help the other player reach the waypoint thereby hitting the checkpoint quickly. On the other hand, if the other player is on any opposing teams, then the ego wants to maximize the squared distance to hinder its opponent's progress towards the checkpoint. This component is calculated by the following function of the waypoint estimated target waypoint $\Bar{\psi}^j$ and a weight parameter $\rho$:
\begin{equation} \label{eq:lqng_obj2}
    \phi^i(\Bar{\psi}^j, \rho) = \sum_{t = 1}^{\Bar{\delta}} \rho((x^j_{t} - \Bar{\psi}^j_x)^2 + (y^j_{t} - \Bar{\psi}^j_y)^2)
\end{equation}

We drop all of the constraints with the exception of collision avoidance, and it is incorporated as the third component and penalty term in the objective where the distance to all other players should be maximized. This term is calculated by the following function of the opponent's position $(x^j_t, y^j_t)$ and a weight parameter $\rho$:
\begin{equation} \label{eq:lqng_obj3}
    \chi^i(x^j_t, y^j_t, \rho) = \sum_{t = 1}^{\Bar{\delta}} \rho((x^j_{t} - x^i_{t})^2 + (y^j_{t} - y^i_{t})^2)
\end{equation}

The final quadratic objective for a player  $i$ on team $\mu$ aggregates \eqref{eq:lqng_obj1}-\eqref{eq:lqng_obj3} using weight multipliers ($\rho_i$) to place varying emphasis on the components as follows:

\begin{small}
\begin{multline} \label{eq:lqng_obj}
    \min_{a^i_{1}, e^i_{1}, ..., a^i_{\Bar{\delta}}, e^i_{\Bar{\delta}}}
    \upsilon^i(\rho_1, \rho_2,\rho_3)
    + \sum_{j \in \{\mu \setminus \{i\}\}}  (\phi^i(\Bar{\psi}^j, \rho_4)) \\
     -\sum_{j \in \{N \setminus \mu\}}  (\phi^i(\Bar{\psi}^j, \rho_5))
    - \sum_{j \in \{N \setminus \{i\}\}} (\chi^i(x^j_t, y^j_t, \rho_6))
\end{multline}
\end{small}

Finally, the dynamics are time invariant and linearized around initial state ($x_{t_0}$, $y_{t_0}$, $v_{t_0}$, $\theta_{t_0}$) for all players $j \in N$:

\begin{small}
\begin{multline} \label{eq:lqng_dyn}
\begin{bmatrix}
x^j_{t+1} \\
y^j_{t+1} \\
v^j_{t+1} \\
\theta^j_{t+1} \\
\end{bmatrix} = 
\begin{bmatrix} 
	1 & 0 & \cos(\theta^j_{t_0})\Delta t & -v^j_{t_0}\sin(\theta^j_{t_0})\Delta t\\
	0 & 1 & \sin(\theta^j_{t_0})\Delta t & v^j_{t_0}\cos(\theta^j_{t_0})\Delta t\\
	0 & 0 & 1 & 0\\
	0 & 0 & 0 & 1\\
	\end{bmatrix}
\begin{bmatrix}
x^j_{t} \\
y^j_{t} \\
v^j_{t} \\
\theta^m_{t} \\
\end{bmatrix} \\ +
\begin{bmatrix} 
	0 & 0 \\
	0 & 0 \\
	\Delta t & 0 \\
	0 & \Delta t \\
	\end{bmatrix}
	\begin{bmatrix} 
	a^j_t  \\
	e^j_t \\
	\end{bmatrix}
\end{multline}
\end{small}

\section{Experiments}
The high-level planner is paired with each of the two low-level planners discussed. We refer to our two hierarchical design variants as MCTS-RL and MCTS-LQNG. 
\subsection{Baseline Controllers}
To measure the importance of our design innovations, we also consider three baseline controllers to resemble the other methods developed in prior works.  

\subsubsection{End-to-End Multi-Agent Reinforcement Learning}
The end-to-end MARL controller, referred to as ``E2E," represents the pure learning-based methods such as that of \cite{sonyai}. This controller has a similar reward/penalty structure as our low-level controller, but its observation structure is slightly different. Instead of observing the sequence of upcoming states as calculated by a high-level planner, E2E only receives the subsequence of locations from $\{c_i\}_{i=1}^{\tau}$ that denote the center of the track near the agent. As a result, it is fully up to its neural networks to learn how to plan strategic and safe moves. 

\subsubsection{Fixed Trajectory Linear-Quadratic Nash Game}
The fixed trajectory LQNG controller, referred to as ``Fixed-LQNG," uses the same LQNG low-level planner as our hierarchical variant, but it tracks a fixed trajectory around the track instead of using a dynamic high-level planner such as our discrete game. This fixed trajectory is a racing line that is computed offline for a specific track using its geometry and parameters of the vehicle as seen in prior works \cite{Vazquez2020, Stahl2019_2}. Furthermore, in the prior works, the method was only applied to single agent racing scenarios, whereas we use the game-theoretic LQNG controller and apply it to multi-agent racing.

\subsubsection{Fixed Trajectory Multi-Agent Reinforcement Learning}
The fixed trajectory MARL controller, referred to as ``Fixed-RL," is a learning-based counterpart to Fixed-LQNG. The online control inputs are computed using a deep RL policy trained to track precomputed checkpoints that are fixed prior to the race.  

\subsection{Experimental Setup}
\begin{figure*}
  \centering
  \includegraphics[width=0.98\textwidth]{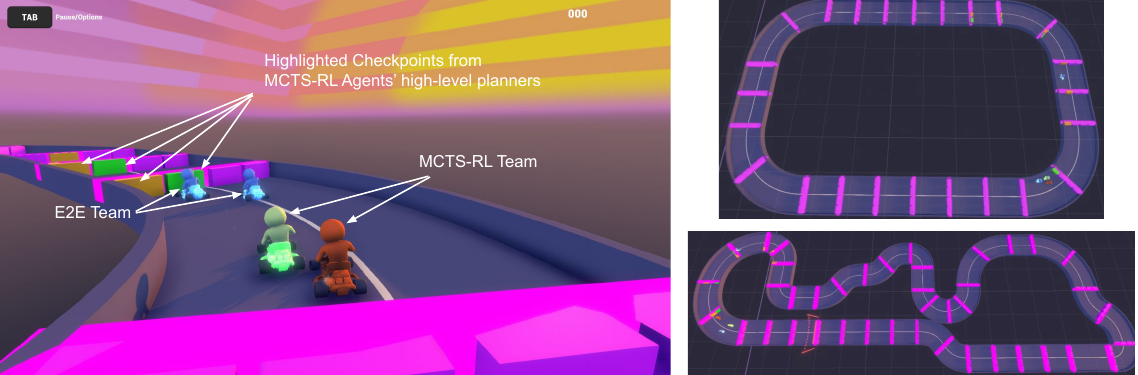}
  \caption{Kart racing environment from an MCTS-RL racer's perspective during a race against an E2E team on the oval track (left). The purple boxes visualize the lanes across checkpoints along the track, and the highlighted green and orange boxes show planned waypoints determined by the hierarchical controllers. We also show a bird's eye view of the oval track (right-top) and complex track (right-bottom) used in our training and experiments.}
  \label{fig:experiment_tracks}
\end{figure*}

Our controllers are implemented\footnote{Code: \codeurl} in the Unity Game Engine. Screenshots of the simulation environment are shown in Figure \ref{fig:experiment_tracks}. We extend the Karting Microgame template \cite{microkarting} provided by Unity. The kart physics from the template is adapted to include cornering limitations and tire wear percentage. Tire wear is modeled as an exponential decay curve that is a function of the accumulated angular velocity endured by the kart. This model captures the concept of losing grip as the tire is subjected to increased lateral loads. Multi-agent support is also added to the provided template in order to race the various autonomous controllers against each other or human players. The high-level planners run at \SI{1}{\hertz}, and low-level planners run at \SI{15}-\SI{50}{\hertz} depending on the number of nearby opponents. The time horizon $\Bar{\delta}$ is set to \SI{0.06}{\second} for the LQNG planner. See the \hyperref[sec:app]{Appendix} for more details regarding the reward functions and training setup for our RL-based agents.

Our experiments include 2v2 team racing on a basic oval track (which the learning-based agents were trained on) and a more complex track (which they were not trained on) shown in Figure \ref{fig:experiment_tracks}. Specifically, the complex track involves challenging track geometry with turns whose radii change along the curves, tight U-turns, and turns in both directions. To be successful, the optimal racing strategy requires some understanding of the shape of the track along a sequence of multiple turns. Each team is composed of two players both using one of the five types of implemented controllers, MCTS-RL, MCTS-LQNG, E2E, Fixed-LQNG, and Fixed-RL, to construct five total teams. Every pair of teams competes head-to-head in 48 races on both tracks. The dynamical parameters of each player's vehicle are identical. The only difference in their initial states is the lane in which they start and the initial checkpoint. Two of the players start \SI{10}{\meter} in front of the other pair resembling the starting grid seen in real-life racing. In order to maintain fairness with respect to starting closer to the optimal racing line or ahead of others, we rotate through each of the six unique ways to place each team on the four possible starting positions.
\subsection{Results} \label{sec:results}
Our experiments seek to reinforce the importance of hierarchical game-theoretic reasoning and study its scalability to challenging problems with strategies requiring decentralized coordination and long-term planning. In our previous work \cite{thakkarprior}, we show that the hierarchical game-theoretic controllers clearly outperform their baselines and exhibit realistic racing maneuvers to overtake and defend in head-to-head scenarios. We also showed how staying close to the plan generated by the high-level tactical planner resulted in better performance but had diminishing returns. Now, we are interested in observing maneuvers where teammates use tactical positioning to help pass or defend against the opposing team, which is also commonly observed in real-life racing. We are also interested in seeing whether the same relationship holds regarding performance with respect to the distance and difference to the high-level plan. 

To obtain a holistic comparison across all of the controllers, we count the number of wins (i.e. 1st place finishes), average collisions-at-fault per race, average illegal lane changes per race, and a safety score (a sum of the prior two metrics). To evaluate team-based performance, we assign points to each of the four finishing positions, $[10, 7.5, 6, 4]$ and $0$ for not finishing the race. The points are summed at the end of the race for each team. To measure the effectiveness of our high-level tactical planner, we also measure average target lane distance and average target velocity difference, which evaluate to the distance and difference in velocity at each checkpoint. This pair of metrics is only collected for MCTS-RL, MCTS-LQNG, and E2E controllers. Though the E2E agents do not rely on the high-level tactical planner, we calculate this metric by running an identically configured planner with respect to the hierarchical agents to see what the target lanes and velocities would be calculated in the E2E agents' perspectives and compare them the E2E controllers' actual decisions. 

Note that the safety score and its component metrics are directly evaluated based on violations to the constraints, Equations \eqref{eq:gen_coll_avoid} and \eqref{eq:gen_lane_lim}, in the original formulation related to these rules. On the other hand, the wins and team points metrics are not a direct measure of the objective function in Equation \eqref{eq:gen_obj}. This is because the objective function is actually an approximation of what it means to win a race. The objective effectively models maximizing performance relative to the average opponent while winning implies being ahead of all opponents. Though the objective is an approximation, we still use wins and team points to compare our controllers because those are the metrics that are ultimately used to measure performance in real-life.

Lastly, we also provide a video\footnote{Video: \vidurl\label{ft:vid}} demonstrating our controllers in action. 

\begin{figure*}
  \centering
  \begin{minipage}{0.5\textwidth}
  \centering
\includegraphics[width=.95\textwidth]{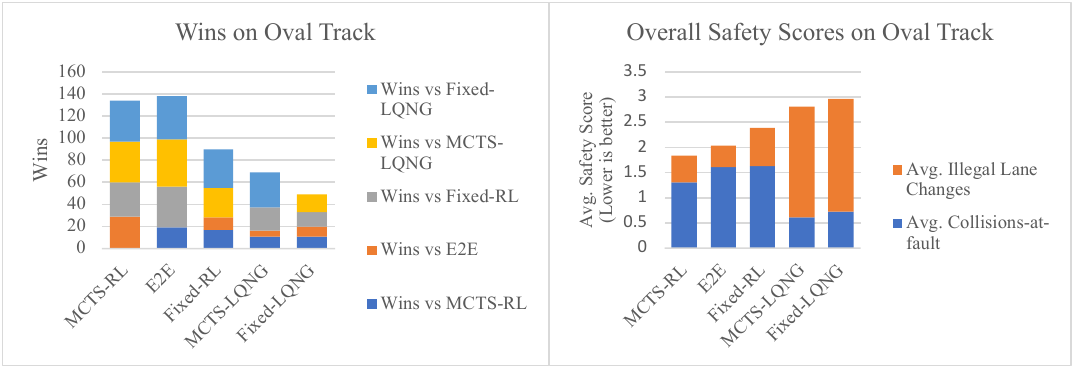}
  \caption{Results racing simulations on the oval track.}
  \label{fig:team_results_oval}
  \end{minipage}\hfill
  \begin{minipage}{0.5\textwidth}  
    \centering 
    \includegraphics[width=.95\textwidth]{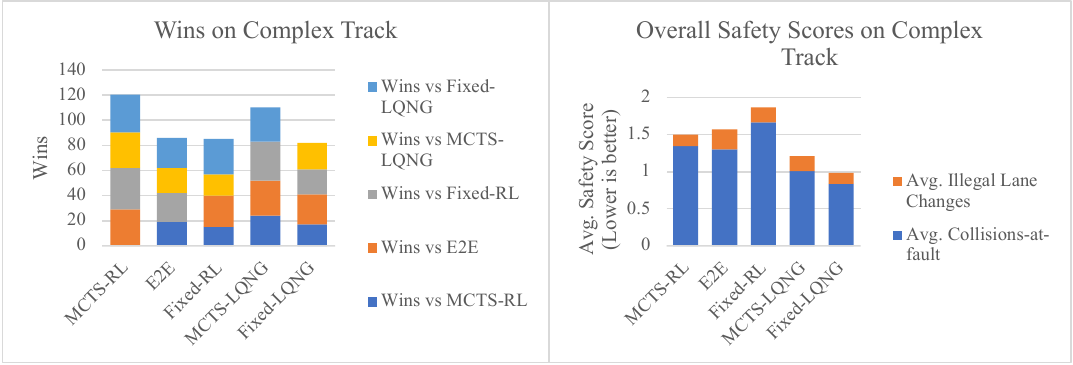}
  \caption{Results racing simulations on the complex track.}
  \label{fig:team_results_complex}
  \end{minipage}
\end{figure*}

\begin{table}[!ht]
    \centering
    \begin{tabular}{|l|l|p{2cm}|p{1.75cm}|}
    \hline
        \textbf{Controller Type} & \textbf{Wins} & \textbf{Average Points Per Race} & \textbf{Average Safety Score} \\ \hline
        MCTS-RL & 254 & 14.93 & 1.61  \\ \hline
        E2E & 224 & 13.20 & 1.67  \\ \hline
        Fixed-RL & 175 & 13.51 & 2.14  \\ \hline
        MCTS-LQR & 175 & 12.89 & 1.91  \\ \hline
        Fixed-LQR & 132 & 12.56 & 1.78 \\ \hline
    \end{tabular}
    \caption[Aggregated team-based racing results.]{Aggregated team-based racing results over 384 total races across both tracks.}
    \label{tab:team_aggr_results}
\end{table}
\begin{table}[!ht]
    \centering
    \begin{tabular}{|l|p{1.75cm}|p{2.25cm}|}
    \hline
        \textbf{Agent Type} & \textbf{Average Lane Distance (\si{\meter})} & \textbf{Average Velocity Difference (\si{\meter\per\second})} \\ \hline
        MCTS-RL & 1.132 & 1.50 \\ \hline
        E2E & 1.564 & 1.529  \\ \hline
        MCTS-LQNG & 0.435 & 0.312  \\ \hline
    \end{tabular}
    \caption{Average distance and difference between the high-level planner's target lane and velocity and the true position and velocity at each checkpoint, respectively.}
    \label{tab:tldv_data}
\end{table}
Based on the plots in Figures \ref{fig:team_results_oval}-\ref{fig:team_results_complex} Tables \ref{tab:team_aggr_results}-\ref{tab:tldv_data}, we conclude the following key points:

\begin{enumerate}[wide, labelindent=0pt, font=\bfseries]
\item \textbf{The proposed hierarchical controllers outperform their respective baselines in team-based racing.} 

The results amongst MCTS-RL, Fixed-RL, and E2E continue to show the effectiveness of our hierarchical structure. Again, all of the MARL-based agents were trained only on the oval track, but MCTS-RL leads in all of the key metrics. While MCTS-RL has more wins overall, the difference in the number of wins is not as high as the head-to-head case in \cite{thakkarprior}. However, the essential metric of interest in this study is average points per race, which evaluates team-based performance. MCTS-RL maintains a considerable difference in terms of average points per race compared to the baselines. The higher points per race implies that even if MCTS-RL is not able to finish first, it collaborates more effectively to produce better results as a team. 

Next, comparing just the baselines, we notice that Fixed-RL is worse in terms of wins and safety score compared to E2E. Recall that the Fixed-RL controller simply follows a fixed optimal racing line. While such a strategy might be successful in the head-to-head case where there is only one opponent to consider, in the cooperative racing scenario, it is imperative for players to consider alternative racing lines especially if one's teammate is already following a specific line. As a result, Fixed-RL often had collisions with its own teammate as both players competed over the same space. In those situations, one or both of the Fixed-RL teammates sometimes lost a position. However, once they were separated far enough after recovering from the collision, both of the agents on the Fixed-RL team could drive fast enough to at least maintain their new positions or sometimes independently overtake its opponents, which is reflected in its higher points-per-race score compared to E2E. This pattern implies that hierarchical reasoning is important to being successful but is not necessarily enough. To be the most successful, game-theoretic hierarchical reasoning, e.g. using MCTS for high-level planning, should be used to allow teammates to predict each other's plans and work together effectively. 

Additionally, without a hierarchical structure, it is easy for a MARL-based controller to overfit. By delegating the primary responsibility for game-theoretic reasoning to the high-level planner, the low-level planner's objective in MCTS-RL is primarily focused on reaching the target waypoints, with less emphasis on tactical reasoning and consideration of the rules. In contrast, E2E is required to encode both tactical planning and the rules in its training, which can and likely did lead to overfitting and difficulty generalizing beyond the training environment. This is reflected in the results, which show that MCTS-RL had many more wins than E2E on the complex track, but slightly fewer the oval track.

Finally, we compare MCTS-LQNG and Fixed-LQNG. Both LQNG agents have similar safety scores. However, MCTS-LQNG still has 33\% more wins and a better points-per-race metric overall. Again, the main drawback with the fixed trajectory tracking agents is that they do not consider alternative racing lines. While in the head-to-head case \cite{thakkarprior} considering alternative lines might not be as important, it becomes considerably more vital to success in multi-agent multi-team racing. 

\item \textbf{Tracking the high-level tactical planner's waypoints results in better overall and team-based performance.}

When we compare the target lane distance and target velocity difference metrics in Table \ref{tab:tldv_data}, we can see the impact of our high-level tactical planner. While the aggregated results indicate the E2E outperformed MCTS-LQNG, most of its successes were on the oval track, which it was likely due to overfitting as discussed in the previous point. When challenged with the complex track, both MCTS-based agents outperformed E2E while also staying closer to the plan generated by by the high-level tactical planner compared to E2E. However, tracking the high-level plan seems to have diminishing returns because MCTS-LQNG performed worse than MCTS-RL although it always stayed close to the generated plan. In general, these results indicate that the high-level tactical planner yields good strategical plans, but there exist cases where blindly following them is not the best choice. The trained, RL-based low-level planner of MCTS-RL allows it to smooth out these situations and know to ignore obviously unacceptable plans. This relates to the shortcoming in our design that the high and low-level planners run independently and concurrently. And because there are no theoretical guarantees about the behavior at either level, we cannot take advantage of having a reliable expectation of how each planner in the controller might behave. As a result, the low-level planner must also be able to reason strategically, which the LQNG struggles with. In the next point, we further compare the performance of RL and LQNG as low-level planners. 

\item \textbf{MARL performs better than LQNG as a low-level planner.}  

The MARL-based agents perform generally better than the LQNG-based agents in terms of our key metrics. However, the difference in their performance is smaller compared to the head-to-head experiments in our previous work \cite{thakkarprior} where the MARL-based agents are considerably better than the LQNG-based counterparts. For example, in the complex track, both the LQNG-based agents have better safety scores than their MARL-based counterparts. However, in the oval track, the MARL-based agents have significantly better safety scores due to the number of illegal lane changes by the LQNG-based agents. his result is likely due to the conservative tuning of the LQNG-based controllers for collision avoidance, which results in fewer collisions-at-fault but also forces them to change lanes more often. Furthermore, it also results in the LQNG-based agents often conceding in close battles and thereby losing races because of the high cost in the planning objective of driving near another player even if there is no collision. Despite that, MCTS-RL has just 45\% more wins in the team-based experiments compared to the 80\% more wins it has against MCTS-LQNG in \cite{thakkarprior}. For the fixed trajectory agents, this gap drops from 250\% to 33\%. Nonetheless, when we consider our primary metric evaluating team-based performance, points-per-race, both MARL-based variants are clearly better than the LQNG-based variants. When all of the results are aggregated across both tracks, all of the metrics are still in favor of using the MARL-based agents because they are generally more robust to nuances of the many possibilities of situations that arise. On the other hand, our LQNG formulation has a mixture of concave and convex components in the objective function, is only linearized around the initial state, and uses short horizons, so our cost surface is sometimes unreliable degrading the resulting behavior.

\item \textbf{MCTS-RL outperforms all other implemented controllers and exhibits teamwork tactics resembling real-life experts.}  

The MCTS-RL team records a win rate of over 66\% of the 384 races it participated in across both tracks, the best overall safety score, and the highest average points per race. The MCTS high-level planner provided the agents a series of waypoints allowing it to make decisions in complex tactical situations where there is a mix of both competitive and cooperative objectives. The MARL-based low-level planner provided robustness to adapt to the multitudes of scenarios that play out. Although the players do not communicate or explicitly coordinate strategies, they still produce cooperative behaviors that improve their overall performance as a team. 

We also observe our control structure execute plans that resemble those performed by expert human drivers. For example, Figure \ref{fig:team_mctsrl:overtake} demonstrates how the two high-level planners of each MCTS-RL agent developed a strategy to perform a pincer like maneuver to overtake an opponent. Both agents from the MCTS-RL team approached the opponent from either side of the opponent at the same time. The opponent could only defend one of the agents on the MCTS-RL team allowing the other agent on the team to pass. In addition, MCTS-RL is also successful at executing strategic maneuvers as seen in Figure \ref{fig:team_mctsrl:strategy} wherein an agent which is ahead momentarily slows down and blocks an opponent behind to allow for its teammate to pass the opponent. The latter example is also a demonstration of long-term planning, in which the orange agent gives up the short term advantage for the long-term gain of having both itself and its teammate ahead of the opponent. Both of these tactics resemble strategies of expert human drivers in real head-to-head racing. The video referenced in Footnote \ref{ft:vid} also demonstrates additional examples of strategical behaviors resembling real-life racing including our hierarchical agent defending against multiple opponents. 
 \end{enumerate}
 
 \begin{figure}
  \centering
  \includegraphics[width=0.48\textwidth]{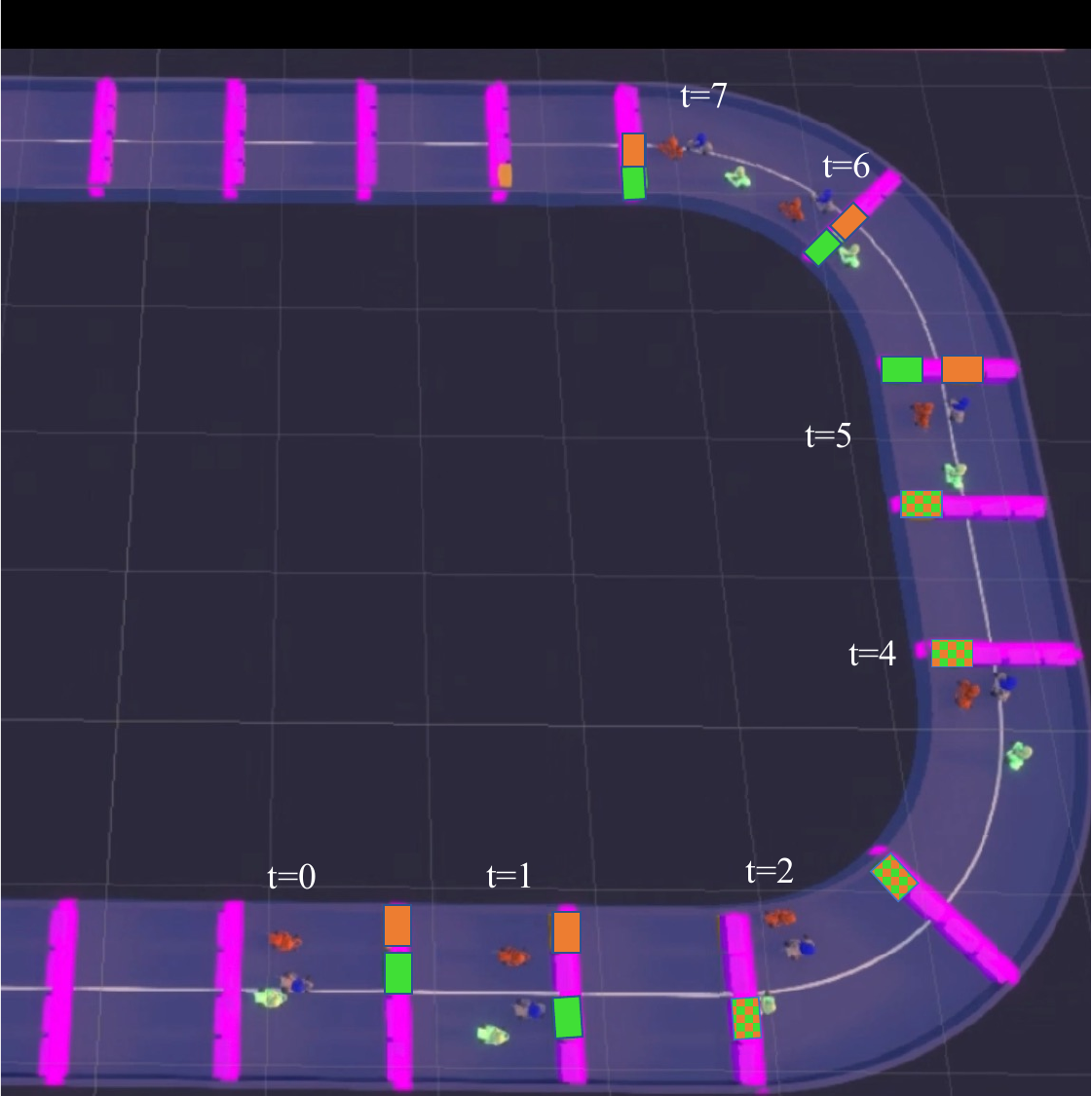}
  \caption [Overtaking maneuver by team of MCTS-RL agents.] {An overtaking maneuver executed a team MCTS-RL agents (green and orange) against the Fixed-RL agent (blue) on the oval track. From $t=0$ to $t=1$, the MCTS-RL agents split and attack the Fixed-RL agent from both sides. The Fixed-RL agent attempts to defend the green MCTS-RL agent on its right allowing the orange MCTS-RL agent to overtake on its left from $t=2$ to $t=6$. The green and orange boxes along each checkpoint highlight the long-term plans calculated by the MCTS planners of each of the MCTS-RL agents, respectively. The checkered boxes indicate a shared checkpoint in their plans.}
  \label{fig:team_mctsrl:overtake}
\end{figure}

\begin{figure*}
 \centering
  \includegraphics[width=\textwidth]{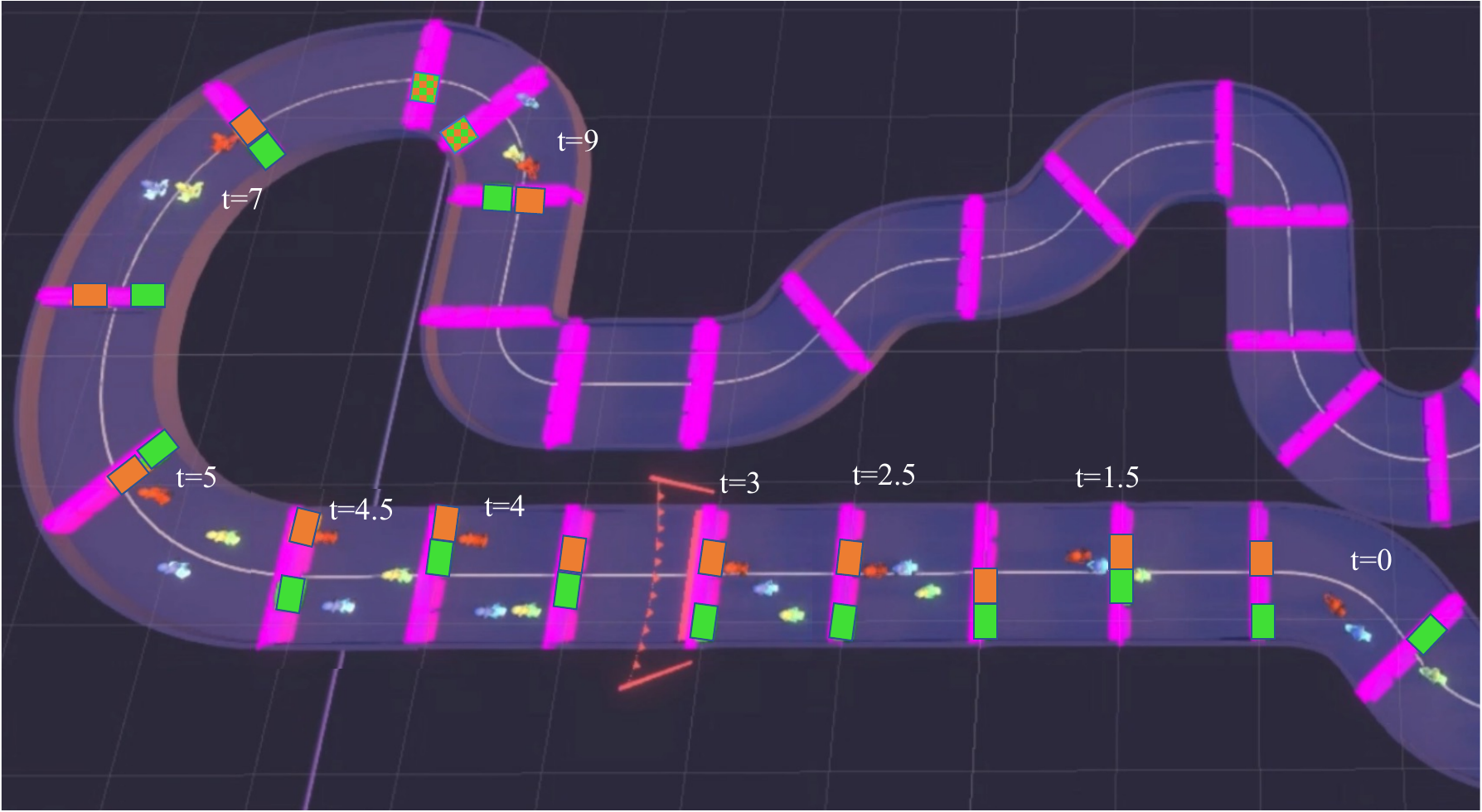}
  \caption[Block to overtake maneuver executed by team of MCTS-RL agents.]{A tactical maneuver executed by the MCTS-RL team (green and orange) against an E2E agent (blue) on the complex track.  Before reaching the turn, the green MCTS-RL's high-level planner calculates a trajectory suggesting to switch lanes to the left for first part of the upcoming straight and block the E2E agent forcing it to slow down and evade further to the left of the track ($t=0$ to $t=3$). This blocking move allows the orange MCTS-RL to plan to take advantage of the opponent's disruption and quickly switch to the right for the inside line of the upcoming turn ( $t=3$ to $t=5$). Eventually, the orange MCTS-RL agent completes the overtake by $t=9$. The green and orange boxes along each checkpoint highlight the long-term plan calculated by MCTS planners of each of the MCTS-RL agents, respectively. The checkered boxes indicate a shared checkpoint in their plans.}
  \label{fig:team_mctsrl:strategy}
\end{figure*}
\section{Conclusion}
We developed a hierarchical controller for cooperative team-based competitive autonomous racing. Our controller outperforms optimization-based and learning-based methods by approximating the complex formulation of the team-based racing game with realistic collision avoidance and lane change constraints.The high-level planner produces long-term trajectories that satisfy these rules, allowing the low-level controllers to focus on tracking the high-level plan and avoiding much of the complexity of the original formulation. Overall, the results indicate that our hierarchical controllers scale to the additional complexities of team-based racing and considering more than two players. They also exhibited maneuvers resembling those performed by expert human drivers such as blocking to aid teammates overtaking chances, pincer-like overtaking moves, and defending against multiple opponents. 

Future extensions of this work should introduce additional high-level and low-level planners. Examples of additional low-level controllers include time-varying linear-quadratic approximations or other nonlinear simplified formulations of the original formulation using iterative best response. With a larger collection of control options, one might investigate policy-switching hierarchical controllers where we switch between the different high and low-level controllers depending on the situation in the game and taking advantage of their strengths.

Lastly, our hierarchical control design can be extended to other multi-agent systems applications where there exist complex rules such as energy grid systems or air traffic control. Constructing a discrete high-level game allows for natural encoding of the complex constraints, often involving discrete components, to find an approximate solution that can warm start a more precise low-level game-theoretic controller.

\bibliographystyle{IEEEtran}
\bibliography{IEEEabrv,bibfile}

\begin{thebibliography}{10}
\providecommand{\url}[1]{#1}
\csname url@samestyle\endcsname
\providecommand{\newblock}{\relax}
\providecommand{\bibinfo}[2]{#2}
\providecommand{\BIBentrySTDinterwordspacing}{\spaceskip=0pt\relax}
\providecommand{\BIBentryALTinterwordstretchfactor}{4}
\providecommand{\BIBentryALTinterwordspacing}{\spaceskip=\fontdimen2\font plus
\BIBentryALTinterwordstretchfactor\fontdimen3\font minus \fontdimen4\font\relax}
\providecommand{\BIBforeignlanguage}[2]{{%
\expandafter\ifx\csname l@#1\endcsname\relax
\typeout{** WARNING: IEEEtran.bst: No hyphenation pattern has been}%
\typeout{** loaded for the language `#1'. Using the pattern for}%
\typeout{** the default language instead.}%
\else
\language=\csname l@#1\endcsname
\fi
#2}}
\providecommand{\BIBdecl}{\relax}
\BIBdecl

\bibitem{adlit}
A.~Faisal, M.~Kamruzzaman, T.~Yigitcanlar, and G.~Currie, ``Understanding autonomous vehicles,'' \emph{Journal of transport and land use}, vol.~12, no.~1, pp. 45--72, 2019.

\bibitem{litreview}
J.~Betz, H.~Zheng, A.~Liniger, U.~Rosolia, P.~Karle, M.~Behl, V.~Krovi, and R.~Mangharam, ``Autonomous vehicles on the edge: A survey on autonomous vehicle racing,'' \emph{IEEE Open Journal of Intelligent Transportation Systems}, vol.~3, pp. 458--488, 2022.

\bibitem{racingadvances}
S.~Edelstein, ``The technologies your car inherited from race cars,'' \url{https://www.digitaltrends.com/cars/racing-tech-in-your-current-car/}, 2019, accessed: 02-17-2022.

\bibitem{Wang2019}
\BIBentryALTinterwordspacing
M.~Wang, Z.~Wang, J.~Talbot, J.~C. Gerdes, and M.~Schwager, ``Game theoretic planning for self-driving cars in competitive scenarios,'' in \emph{Robotics: Science and Systems XV, University of Freiburg, Freiburg im Breisgau, Germany, June 22-26, 2019}, A.~Bicchi, H.~Kress{-}Gazit, and S.~Hutchinson, Eds., 2019. [Online]. Available: \url{https://doi.org/10.15607/RSS.2019.XV.048}
\BIBentrySTDinterwordspacing

\bibitem{Wang2021}
\BIBentryALTinterwordspacing
------, ``Game-theoretic planning for self-driving cars in multivehicle competitive scenarios,'' \emph{{IEEE} Transactions on Robotics}, pp. 1--13, 2021. [Online]. Available: \url{https://doi.org/10.1109/tro.2020.3047521}
\BIBentrySTDinterwordspacing

\bibitem{Li2021}
\BIBentryALTinterwordspacing
N.~Li, E.~Goubault, L.~Pautet, and S.~Putot, ``Autonomous racecar control in head-to-head competition using mixed-integer quadratic programming,'' in \emph{2021 International Conference on Robotics and Automation (ICRA 2021) - Workshop Opportunities and Challenges With Autonomous Racing}.\hskip 1em plus 0.5em minus 0.4em\relax {IEEE}, 2021. [Online]. Available: \url{https://linklab-uva.github.io/icra-autonomous-racing/contributed_papers/paper2.pdf}
\BIBentrySTDinterwordspacing

\bibitem{He2021}
\BIBentryALTinterwordspacing
S.~He, J.~Zeng, and K.~Sreenath, ``Autonomous racing with multiple vehicles using a parallelized optimization with safety guarantee using control barrier functions,'' \emph{arXiv preprint arXiv:2112.06435}, 2022. [Online]. Available: \url{https://arxiv.org/abs/2112.06435}
\BIBentrySTDinterwordspacing

\bibitem{racingrules}
T.~Martin, ``The guide to road racing, part 8: Passing etiquette,'' \url{https://www.windingroad.com/articles/blogs/the-road-racers-guide-to-passing-etiquette/}, 2020, accessed: 02-17-2022.

\bibitem{thakkarprior}
\BIBentryALTinterwordspacing
R.~S. Thakkar, A.~S. Samyal, D.~Fridovich-Keil, Z.~Xu, and U.~Topcu, ``Hierarchical control for head-to-head autonomous racing,'' \emph{Field Robotics}, vol.~4, pp. 46--69, 2024. [Online]. Available: \url{https://fieldrobotics.net/Field_Robotics/Volume_4_files/Vol4_02.pdf}
\BIBentrySTDinterwordspacing

\bibitem{Xu2021}
J.~Xu, C.~Yan, Y.~Xu, J.~Shi, K.~Sheng, and X.~Xu, ``A hierarchical game theory based demand optimization method for grid-interaction of energy flexible buildings,'' \emph{Frontiers in Energy Research}, p. 500, 2021.

\bibitem{Tomlin1996}
\BIBentryALTinterwordspacing
C.~Tomlin, G.~Pappas, J.~Lygeros, D.~Godbole, S.~Sastry, and G.~Meyer, ``Hybrid control in air traffic management systems1,'' \emph{IFAC Proceedings Volumes}, vol.~29, no.~1, pp. 5512--5517, 1996, 13th World Congress of IFAC, 1996, San Francisco USA, 30 June - 5 July. [Online]. Available: \url{https://www.sciencedirect.com/science/article/pii/S1474667017585596}
\BIBentrySTDinterwordspacing

\bibitem{Hou2016}
J.-H. Hou and T.~Wang, ``The development of a simulated car racing controller based on monte-carlo tree search,'' in \emph{2016 Conference on Technologies and Applications of Artificial Intelligence (TAAI)}, 2016, pp. 104--109.

\bibitem{Vazquez2020}
\BIBentryALTinterwordspacing
J.~L. Vazquez, M.~Bruhlmeier, A.~Liniger, A.~Rupenyan, and J.~Lygeros, ``Optimization-based hierarchical motion planning for autonomous racing,'' in \emph{2020 {IEEE}/{RSJ} International Conference on Intelligent Robots and Systems ({IROS})}.\hskip 1em plus 0.5em minus 0.4em\relax {IEEE}, Oct. 2020. [Online]. Available: \url{https://doi.org/10.1109/iros45743.2020.9341731}
\BIBentrySTDinterwordspacing

\bibitem{Stahl2019_2}
\BIBentryALTinterwordspacing
T.~Stahl, A.~Wischnewski, J.~Betz, and M.~Lienkamp, ``Multilayer graph-based trajectory planning for race vehicles in dynamic scenarios,'' in \emph{2019 {IEEE} Intelligent Transportation Systems Conference ({ITSC})}.\hskip 1em plus 0.5em minus 0.4em\relax {IEEE}, Oct. 2019. [Online]. Available: \url{https://doi.org/10.1109/itsc.2019.8917032}
\BIBentrySTDinterwordspacing

\bibitem{Kabzan2019}
\BIBentryALTinterwordspacing
J.~Kabzan, L.~Hewing, A.~Liniger, and M.~N. Zeilinger, ``Learning-based model predictive control for autonomous racing,'' \emph{{IEEE} Robotics and Automation Letters}, vol.~4, no.~4, pp. 3363--3370, Oct. 2019. [Online]. Available: \url{https://doi.org/10.1109/lra.2019.2926677}
\BIBentrySTDinterwordspacing

\bibitem{Remonda2021}
\BIBentryALTinterwordspacing
A.~Remonda, S.~Krebs, E.~E. Veas, G.~Luzhnica, and R.~Kern, ``Formula {RL:} deep reinforcement learning for autonomous racing using telemetry data,'' \emph{CoRR}, vol. abs/2104.11106, 2021. [Online]. Available: \url{https://arxiv.org/abs/2104.11106}
\BIBentrySTDinterwordspacing

\bibitem{deBruin2018}
\BIBentryALTinterwordspacing
T.~de~Bruin, J.~Kober, K.~Tuyls, and R.~Babuska, ``Integrating state representation learning into deep reinforcement learning,'' \emph{{IEEE} Robotics and Automation Letters}, vol.~3, no.~3, pp. 1394--1401, Jul. 2018. [Online]. Available: \url{https://doi.org/10.1109/lra.2018.2800101}
\BIBentrySTDinterwordspacing

\bibitem{weiss2020}
\BIBentryALTinterwordspacing
T.~Weiss and M.~Behl, ``Deepracing: Parameterized trajectories for autonomous racing,'' \emph{arXiv preprint arXiv:2005.05178}, 2020. [Online]. Available: \url{https://arxiv.org/abs/2005.05178}
\BIBentrySTDinterwordspacing

\bibitem{spica2020real}
R.~Spica, E.~Cristofalo, Z.~Wang, E.~Montijano, and M.~Schwager, ``A real-time game theoretic planner for autonomous two-player drone racing,'' \emph{IEEE Transactions on Robotics}, vol.~36, no.~5, pp. 1389--1403, 2020.

\bibitem{Schwarting2021}
\BIBentryALTinterwordspacing
W.~Schwarting, T.~Seyde, I.~Gilitschenski, L.~Liebenwein, R.~Sander, S.~Karaman, and D.~Rus, ``Deep latent competition: Learning to race using visual control policies in latent space,'' \emph{arXiv preprint arXiv:2102.09812}, 2021. [Online]. Available: \url{https://arxiv.org/abs/2102.09812}
\BIBentrySTDinterwordspacing

\bibitem{Song2021}
\BIBentryALTinterwordspacing
Y.~Song, H.~Lin, E.~Kaufmann, P.~Duerr, and D.~Scaramuzza, ``Autonomous overtaking in gran turismo sport using curriculum reinforcement learning,'' \emph{arXiv preprint arXiv:2103.14666}, 2021. [Online]. Available: \url{https://arxiv.org/abs/2103.14666}
\BIBentrySTDinterwordspacing

\bibitem{sonyai}
P.~R. Wurman, S.~Barrett, K.~Kawamoto, J.~MacGlashan, K.~Subramanian, T.~J. Walsh, R.~Capobianco, A.~Devlic, F.~Eckert, F.~Fuchs \emph{et~al.}, ``Outracing champion gran turismo drivers with deep reinforcement learning,'' \emph{Nature}, vol. 602, no. 7896, pp. 223--228, 2022.

\bibitem{LinigerThesis}
A.~Liniger, ``Path planning and control for autonomous racing,'' Ph.D. dissertation, ETH Z\"{u}rich, 2018.

\bibitem{Fisac2019}
J.~F. Fisac, E.~Bronstein, E.~Stefansson, D.~Sadigh, S.~S. Sastry, and A.~D. Dragan, ``Hierarchical game-theoretic planning for autonomous vehicles,'' in \emph{2019 International Conference on Robotics and Automation (ICRA)}.\hskip 1em plus 0.5em minus 0.4em\relax IEEE, 2019, pp. 9590--9596.

\bibitem{Moghadam2019}
M.~Moghadam and G.~H. Elkaim, ``A hierarchical architecture for sequential decision-making in autonomous driving using deep reinforcement learning,'' \emph{arXiv preprint arXiv:1906.08464}, 2019.

\bibitem{wang2017cooperative}
Y.~Wang, E.~Garcia, D.~Casbeer, and F.~Zhang, \emph{Cooperative control of multi-agent systems: Theory and applications}.\hskip 1em plus 0.5em minus 0.4em\relax John Wiley \& Sons, 2017.

\bibitem{Abdoos2021}
M.~Abdoos, ``A cooperative multiagent system for traffic signal control using game theory and reinforcement learning,'' \emph{IEEE Intelligent Transportation Systems Magazine}, vol.~13, no.~4, pp. 6--16, 2021.

\bibitem{hwang2004cooperative}
K.-S. Hwang, S.-W. Tan, and C.-C. Chen, ``Cooperative strategy based on adaptive q-learning for robot soccer systems,'' \emph{IEEE Transactions on Fuzzy Systems}, vol.~12, no.~4, pp. 569--576, 2004.

\bibitem{Rajamani2011vehicle}
R.~Rajamani, \emph{Vehicle dynamics and control}.\hskip 1em plus 0.5em minus 0.4em\relax Springer Science \& Business Media, 2011.

\bibitem{mcts}
R.~Coulom, ``Efficient selectivity and backup operators in monte-carlo tree search,'' in \emph{International conference on computers and games}.\hskip 1em plus 0.5em minus 0.4em\relax Springer, 2006, pp. 72--83.

\bibitem{basar}
T.~Ba{\c{s}}ar and G.~J. Olsder, \emph{Dynamic noncooperative game theory}.\hskip 1em plus 0.5em minus 0.4em\relax SIAM, 1998.

\bibitem{microkarting}
U.~Technologies, ``Unity technologies karting microgame template,'' \url{https://assetstore.unity.com/packages/templates/karting-microgame-150956}, 2021, accessed: 10-2021.

\bibitem{mlagents}
A.~Juliani, V.-P. Berges, E.~Teng, A.~Cohen, J.~Harper, C.~Elion, C.~Goy, Y.~Gao, H.~Henry, M.~Mattar \emph{et~al.}, ``Unity: A general platform for intelligent agents,'' \emph{arXiv preprint arXiv:1809.02627}, 2018.

\bibitem{poca}
A.~Cohen, E.~Teng, V.-P. Berges, R.-P. Dong, H.~Henry, M.~Mattar, A.~Zook, and S.~Ganguly, ``On the use and misuse of absorbing states in multi-agent reinforcement learning,'' \emph{arXiv preprint arXiv:2111.05992}, 2021.

\end{thebibliography}

\section{Appendix} \label{sec:app}
\subsection{Multi-Agent Reinforcement Learning Controller Reward Structure Details}
We outline the specifics of the reward and penalty calculations in detail for our Multi-Agent Reinforcement Learning (MARL) low-level controller. Recall that the MARL-based agents observations include perfect state information for all players (including $(x, y)$ position, $v$ velocity, lane ID $a$, ``recent'' lane change count $e$, and last passed checkpoint $r$) and 9 LIDAR rays, whose distances we refer to as $I_1, \ldots, I_9$. Furthermore, we also assume players know the overall time elapsed in the game $t$, and the maximum time horizon $T$. 

We list functions $R(\cdot)$ that evaluate the rewards or penalties based on one or more weight parameters denoted by $\omega_i$. Our rewards and penalties are categorized into two types:
\begin{enumerate}
\item For every time step in the environment, we provide the following rewards and penalties:
\begin{itemize}
    \item A reward for driving fast. The reward that scales based on the driving close to the top speed of the kart.
    \begin{equation*}
        R_{\text{speed}}(\omega_1) = \omega_1 \frac{v}{v_{\text{max}}}
    \end{equation*}
    \item A reward for moving towards the next checkpoint $r*$. We use the three-dimensional velocity vector of the agent and take the dot product with the vector between the agent's position and the next checkpoint position.
    \begin{equation*}
        R_{\text{direction}}(\omega_1) = \omega_1 (\langle v_x, v_y \rangle \cdot \langle r^*_x-x , r^*_y - y\rangle)
    \end{equation*}
    \item A penalty for exceeding the lane changing limit. We use an indicator function to determine if the player is in the straight region of the track $\mathcal{S}$ and whether the lane changing limit $L$ is exceeded.
    \begin{equation*}
        R_{\text{swerve}}(\omega_1) = -\omega_1 \mathds{1}_{(x,y) \in \mathcal{S} \wedge e > L}
    \end{equation*}
    \item A penalty for being within $h$ meters of the wall. We use an indicator function $\mathds{1}_{I_j < h \wedge I_j \text{hit wall}}$ that determines if he LIDAR reading is below $h$ and if whether the LIDAR bounced off a player or a wall.
    \begin{equation*}
        R_{\text{wall-hit}}(\omega_1) = -\sum_{j=1}^9 \omega_1 \mathds{1}_{I_j < h \wedge I_j \text{hit wall}}
    \end{equation*}
    \item A penalty for being within $h$ meters of another player. Using a similar indicator function from above, if any LIDAR ray in that set hits another player within a distance $h$, then the original player is penalized for being in collision. In addition, we assume we have a set $\Theta$, which contains the indices of the LIDAR rays that point towards the front of the kart. There is an additional penalty if the LIDAR rays come from the subset $\Theta$ as that indicates some form of rear-end collision where the player would be at fault.
    \begin{small}
    \begin{equation*}
        R_{\text{player-hit}}(\omega_1, \omega_2) = -\sum_{j=1}^9 (\omega_1 \mathds{1}_{I_j < h \wedge I_j \text{hit player}} + \omega_2 \mathds{1}_{j \in \Theta})
    \end{equation*}
    \end{small}
\end{itemize}

\item When a player passes a checkpoint with index $r'$, we provide the following rewards and penalties:
\begin{itemize}
    \item A reward to teach the policy to pass as many checkpoints as possible before other players. The reward is scaled based on the order in which the checkpoint is reached. This reward is also added (with a different weight parameter) to a shared reward value used by the posthumous credit assignment algorithm to incentivize cooperative behavior. 
    \begin{equation*}
        R_{\text{checkpoint base}}(\omega_1) = 
        \begin{cases}
            \omega_1 & \text{if first} \\
            0.75\omega_1 & \text{if second} \\
            0.6\omega_1 & \text{if third} \\
            0.4\omega_1 & \text{if fourth} \\
        \end{cases}
    \end{equation*}
    \item A reward based on the remaining time in the game to incentivize minimizing time between checkpoints. This reward is also added (with a different weight parameter) to a shared reward value used by the posthumous credit assignment algorithm to incentivize cooperative behavior. 
    \begin{equation*}
        R_{\text{checkpoint time}}(\omega_1) = \omega_1\frac{T-t}{T}
    \end{equation*}
    \item A reward for being closer to the target lane $a'$ and velocity $v'$ for the passed checkpoint.
    \begin{align*}
        R_{\text{checkpoint target}}(\omega_1, \omega_2) = & \frac{\omega_1}{1.3^{|a-a'|\sqrt{(a'_x-x)^2+(a'_y-y)^2}}} \\
        & + \frac{\omega_2}{1.1^{|v-v'|}}
       \end{align*}
    \item A penalty for driving in reverse. We use an indicator function to determine if checkpoint index $r'$ is less than or equal to $r$ implying the player passed either the same checkpoint or an earlier checkpoint.
    \begin{equation*}
        R_{\text{checkpoint reverse}}(\omega_1) = -\omega_1\mathds{1}_{r' \leq r}
    \end{equation*}
\end{itemize}
\end{enumerate}
\subsection{Multi-Agent Reinforcement Learning Controller Architecture and Training}
We follow an almost identical training procedure as we did in the prequel to this paper \cite{thakkarprior}. We continue use the Unity library known as ML-Agents \cite{mlagents} to train the RL-based agents. However, in order to train the cooperative agents, we use an algorithm, created by the developers of Unity ML-Agents, titled posthumous credit assignment \cite{poca}. This algorithm is an extension of the popular multi-agent reinforcement learning algorithm, counterfactual multi-agent policy gradients, but it modifies how the agents' policies are impacted even when they have reached an absorbing state while other agents who may be on the absorbed agent's team are still alive. In our case, the absorbing state refers to an agent reaching the finish line, i.e. final checkpoint. The training environment consists of eight copies of two sizes of oval tracks. Within each set of tracks, half of the training assumed a clockwise race direction and the other half assumed a counter-clockwise direction. Using two sizes of tracks ensures that the agents learn to make both sharp and wide turns, and using the two race directions allows the agents to learn to make both left and right turns. However, the training is limited to just those track configurations to limit overfitting and evaluate how the various controllers generalize to unknown environments such as the complex track. 

\begin{figure}
\centering
\includegraphics[width=0.48\textwidth]{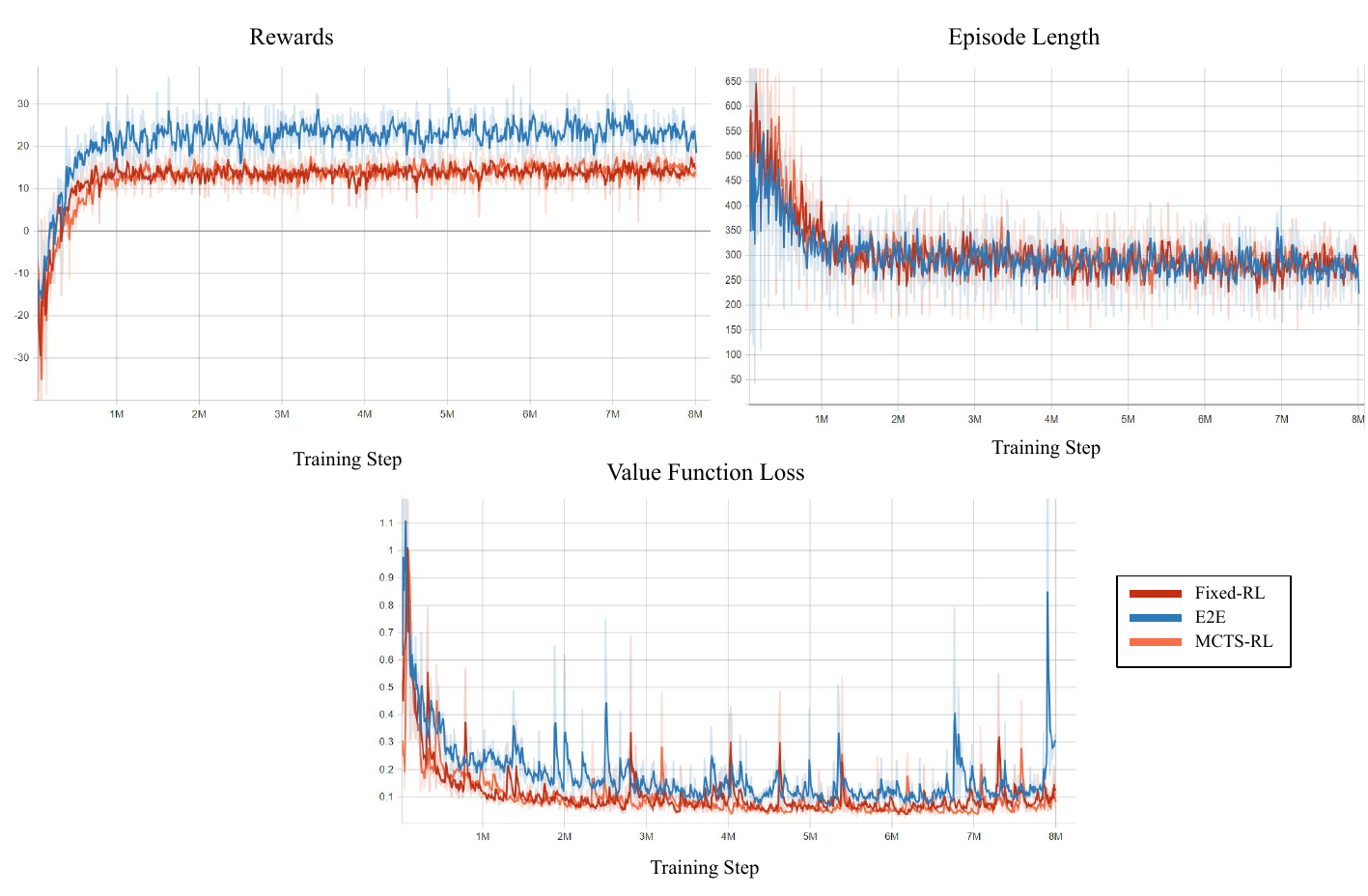}
  \caption{Plots of the rewards (top-left), episode lengths (top-right), and value function losses (bottom) convergence over the training of the RL-based models.}
  \label{fig:train_net}
\end{figure}

\begin{figure}
  \centering
\includegraphics[width=0.45\textwidth]{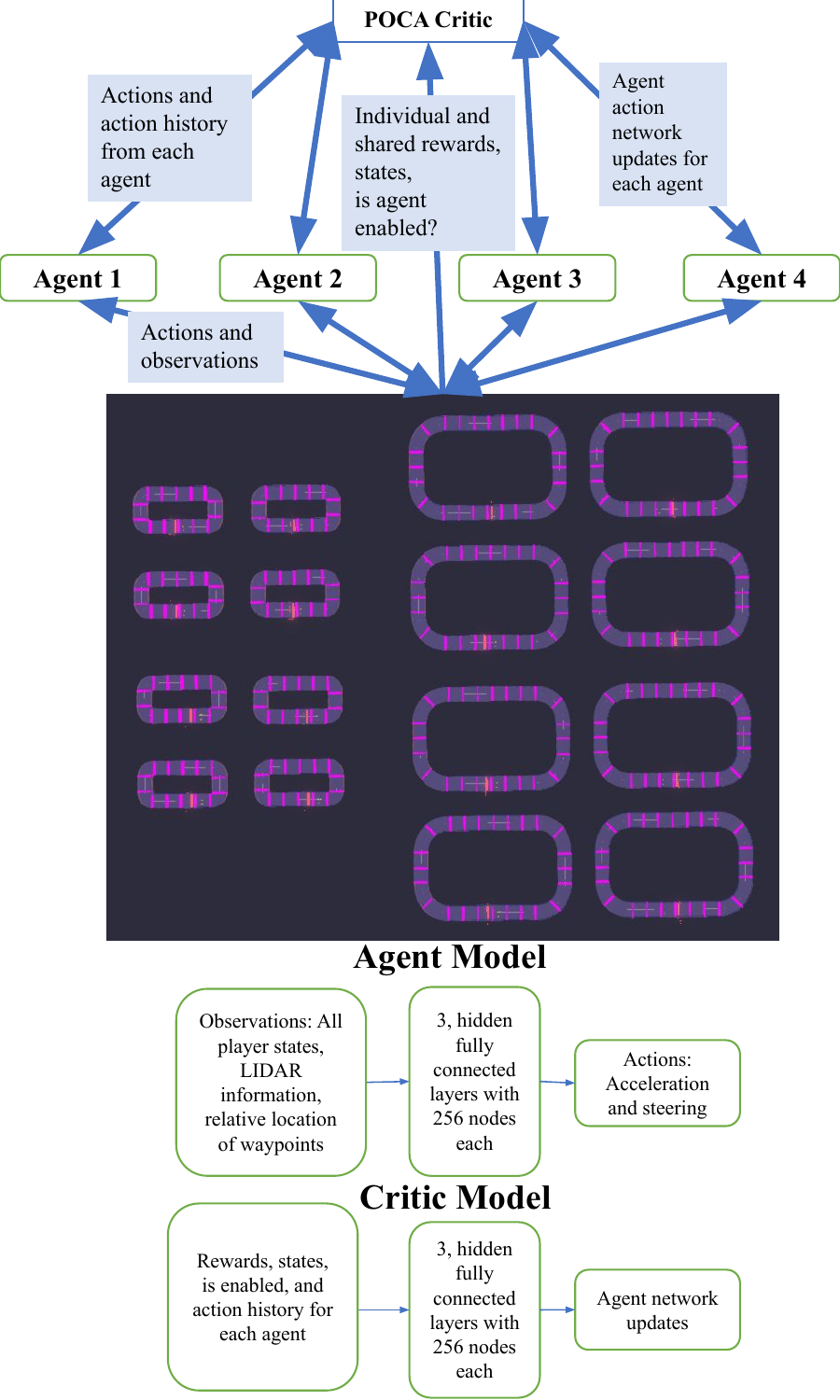}
  \caption{Visualization of the training environment and neural network structures used for all of the RL-based models. The only difference in the training of the three agents is the weights of the reward structures.}
  \label{fig:train_mod}
\end{figure}

The agents share model inputs, policy and reward network sizes and structures, and model outputs. The input is a matrix consisting of stacked vectors of previously mentioned observations (own state, LIDAR rays, opponent state, checkpoint progress, etc.). Both the actor and critic networks consist of 3 hidden layers with 256 nodes each. Figure \ref{fig:train_mod} is a visualization of the described training environment, and Figure \ref{fig:train_net} presents the reward, episode length, and value function loss graphs across training showing their convergence. Note that the rewards scale varies amongst the three types of agents because the weights in the reward functions are different. However, all of the agents are trained to 8000000 steps and their rewards stabilized before reaching the step limit as seen in the graph. 

\begin{table}
\centering
\caption{Symbols used in formulations}
\begin{tabular}{p{0.15\linewidth}|p{0.70\linewidth}}  \label{tab:symbols}
Symbol & Value\\ 
\hline
$N$ & Set of players in the game \\
$M$ & Set of mutually exclusive subsets of $N$ representing teams of players \\
$\mathcal{T}$ & Set of steps in the game \\
$\delta$ & Number of steps in the game \\
$\{c_i\}_{i=1}^{\tau}$ & Sequence of checkpoints \\
$C$ & Set of checkpoint indices \\
$x^i_t$     &   Continuous state of player $i$ at time $t$   \\
$u^i_t$      &   Control input of player $i$ at time $t$    \\
$f(x^i_t, u^i_t)$  &  Continuous dynamics of player $i$ \\ 
$r^i_t$      &  Index of last checkpoint passed by player $i$ at time $t$ \\ 
$\gamma^i$     &  Time when player $i$ passed last checkpoint \\     
$\zeta$ & A multiplier to balance emphasis on cooperative vs. selfish objectives. \\
$p(x^i_{t+1}, r^i_t)$  &  Function computing index of last track checkpoint passed  \\     
$q(x^i_{t+1})$   &  Function computing minimum distance to last track checkpoint passed \\
$w$      &  track width \\     
$d(x^i_t, x^j_t)$ & Function computing distance between player i and player j \\
$s_0$ & Minimum distance safety margin if player is not directly behind another \\
$s_1$ & Minimum distance safety margin if player is directly behind another  \\
$\lambda$ & Number of lanes in the track \\
$l^i_t$ & Integer state variable indicating player $i$'s recent lane changes at time $t$ \\
$y(x^i_t)$ & Function evaluating if player i is on a straight or curve \\
$z(x^i_t)$ & Function computing the lane of the track player is in\\
$L$ & Upper bound on the number of recent lane changes a player is allowed \\
$a^i_k$ & Lane ID for player $i$ at checkpoint $k$\\
$l^i_k$ & Recent lane change count for player $i$ at checkpoint $k$\\
$v^i_k$ & Discrete velocity state for player $i$ at checkpoint $k$\\
$e^i_k$ & Discrete tire wear state for player $i$ at checkpoint $k$\\
$t^i_k$ & Discrete time state for player $i$ at checkpoint $k$\\
$w_l$ & Lane width\\
$\upsilon_{k,k+1}$ & Straight line distance between two checkpoints\\
$a_{\text{max}}$ & Maximum lateral acceleration the vehicle can sustain\\
$a_{\text{min}}$ & Lateral acceleration that the vehicle sustain for any tire wear level\\
$v_{\text{max}}$ & Maximum speed of vehicle\\
$a$ & Maximum acceleration of vehicle\\
$b$ & Maximum deceleration of vehicle\\
$L_{\text{straight}}$ & Tire wear factor parameter for straight sections of track\\
$L_{\text{curve}}$ & Tire wear factor parameter for curve sections of track\\
$\hat{\mathcal{T}}$ & Set of steps in the low-level game \\
$\hat{\delta}$ & Shortened horizon \\
$\alpha$      &   Weight parameter in objective emphasizing importance of hitting trajectory waypoints    \\
$\psi^i_c$      &   Waypoint for player $i$ to target when passing checkpoint index $c$ \\
$\eta^i_c$  &  Distance of player $i$'s closest approach to the waypoint $\psi^i_c$ \\ 
$h(x^i_{t}, \psi^i_c)$  &  Function distance of player $i$'s state to waypoint $\psi^i_c$ \\  
\end{tabular}
\end{table}

\newpage
\begin{biography}[{
\includegraphics[width=1in,height=1.25in,clip,keepaspectratio]{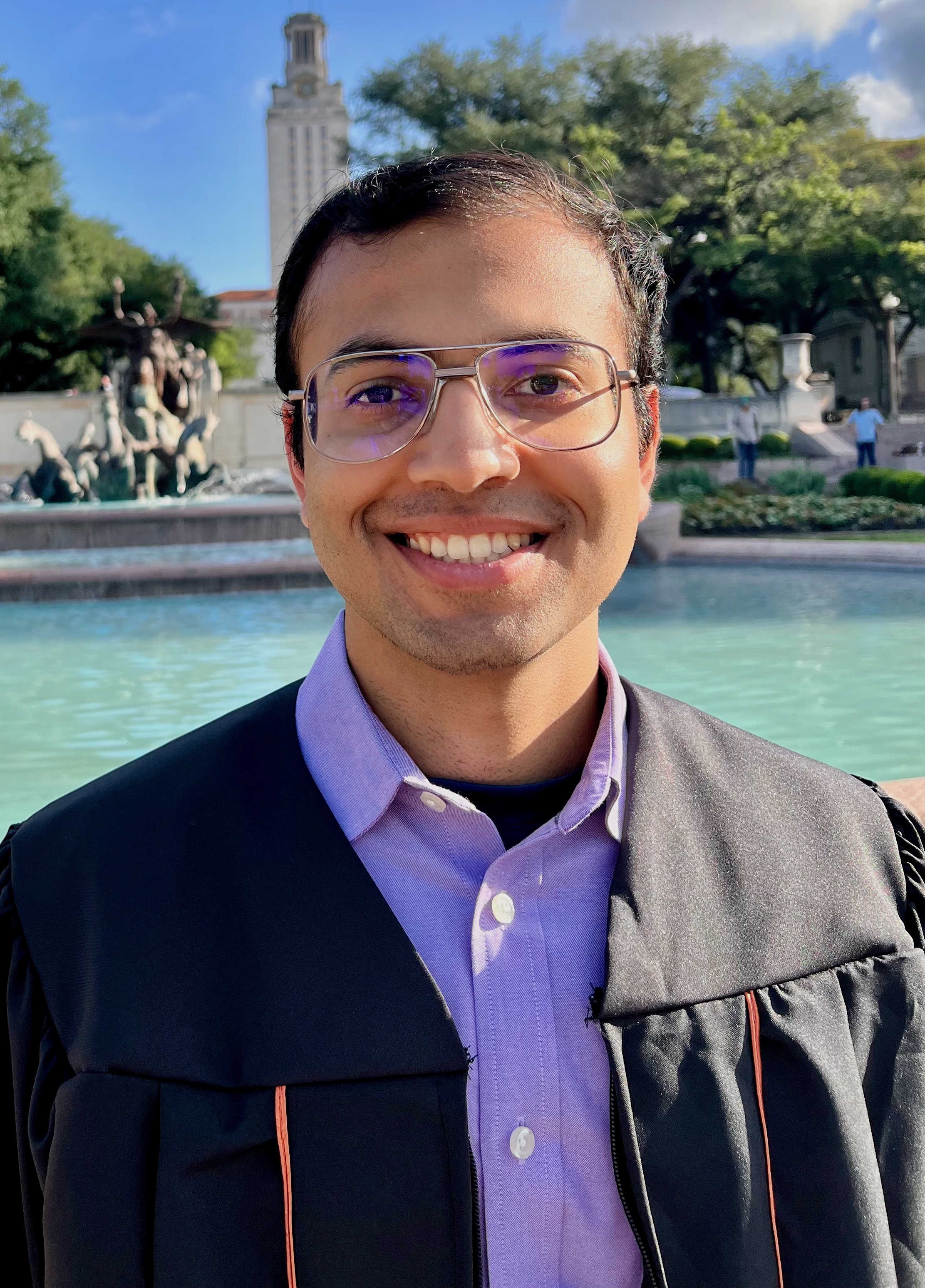}
}]{Rishabh Saumil Thakkar} is an automated trading systems software engineer at Optiver in Chicago, IL. He received an M.S. in Computational Science, Engineering, and Math from the Oden Institute for Computational Engineering and Sciences and a B.S. in Computer Science at the University of Texas at Austin. His interests include autonomous systems, optimization, transportation, and high-performance software engineering.
\end{biography}
\vskip -2\baselineskip plus -1fil
\begin{biography}[{
\includegraphics[width=1in,height=1.25in,clip,keepaspectratio]{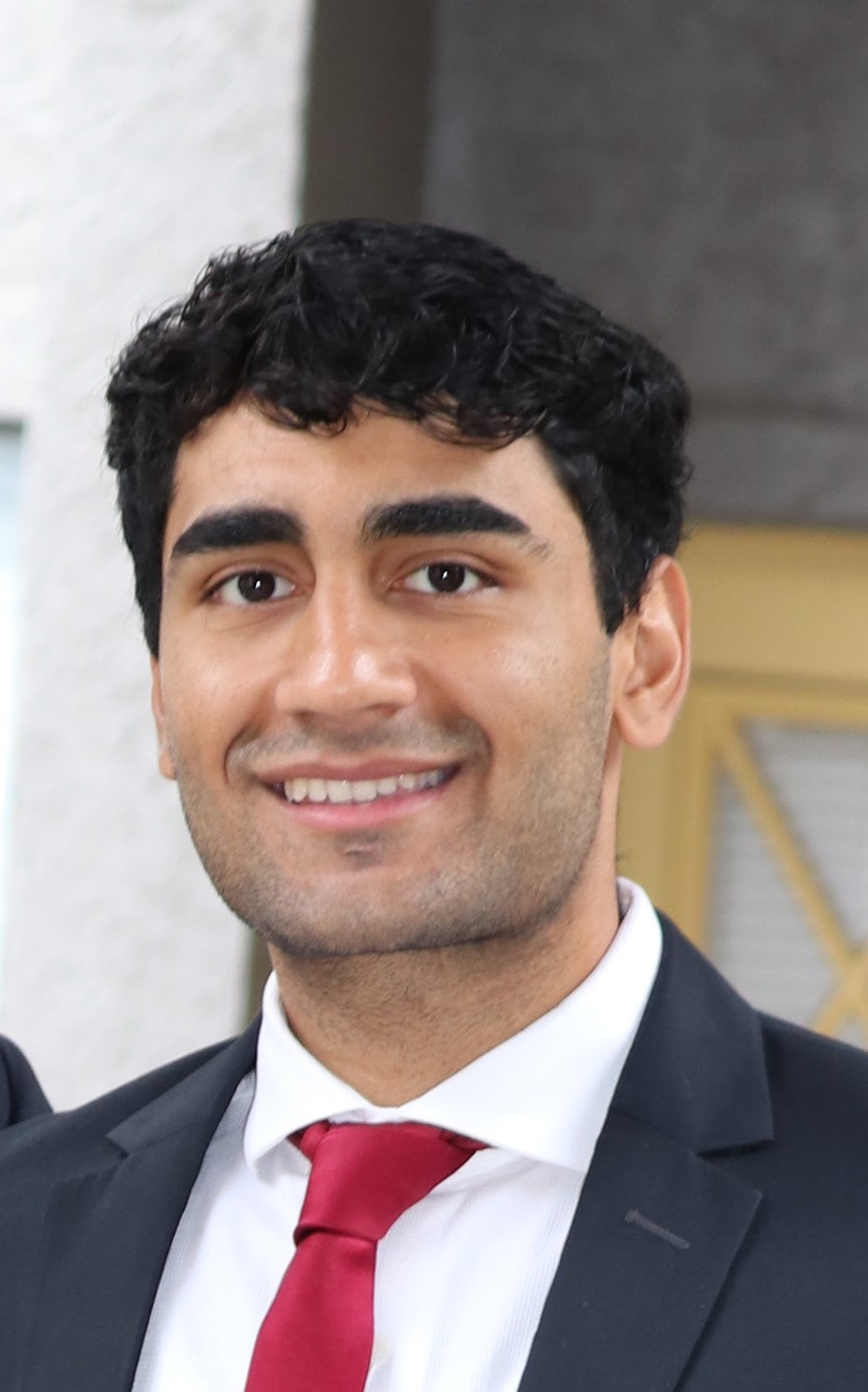}
}]{Aryaman Singh Samyal} is a Machine Learning Research Engineer at Skygrid based in Austin, TX. Holding an M.S. in Aerospace Engineering from UT Austin and a BTech in Aeronautical Engineering from Manipal Institute of Technology, Aryaman specializes in applied ML, focusing on deep learning, reinforcement learning, and autonomous systems. His research aims to enable autonomy in aerospace systems by using learning-based methods.
\end{biography}
\vskip -2\baselineskip plus -1fil
\begin{biography}[{
\includegraphics[width=1in,height=1.25in,clip,keepaspectratio]{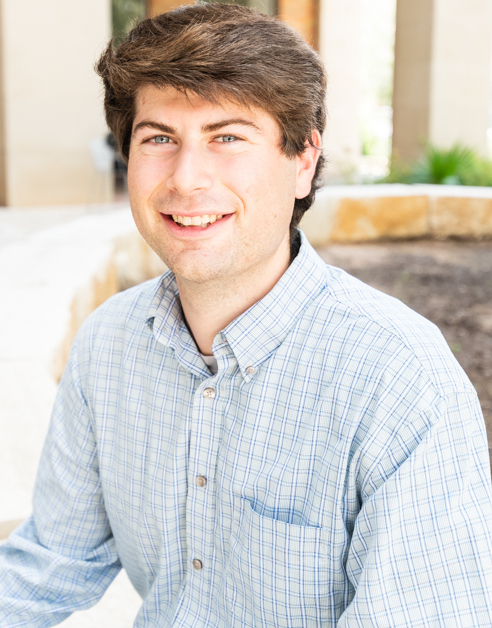}
}]{David Fridovich-Keil} (Member, IEEE) received the B.S.E. degree in Electrical Engineering from Princeton University, and the Ph.D. degree in Electrical Engineering and Computer Sciences from the University of California, Berkeley. He is currently an assistant professor in the Department of Aerospace Engineering and Engineering Mechanics at the University of Texas at Austin. His research interests include optimization, optimal control theory, and game theory. Fridovich-Keil is a recipient of the NSF Graduate Research Fellowship and NSF CAREER Award.
\end{biography}
\vskip -2\baselineskip plus -1fil
\begin{biography}[{
\includegraphics[width=1in,height=1.25in,clip,keepaspectratio]{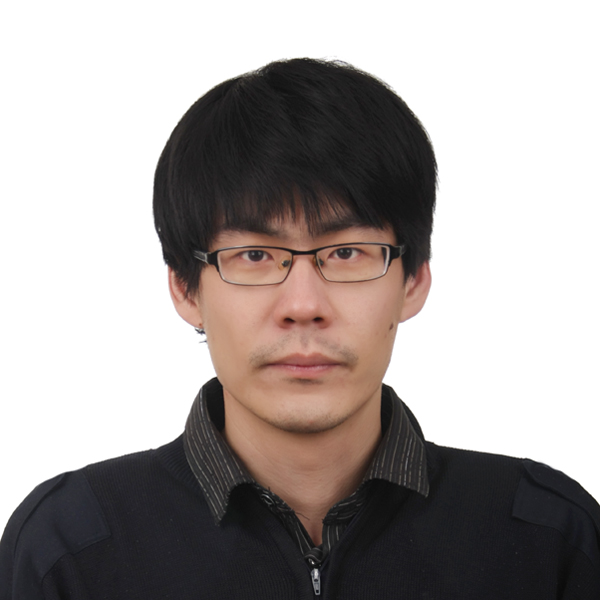}
}]{Zhe Xu} received the B.S. and M.S. degrees in Electrical Engineering from Tianjin University, Tianjin, China, in 2011 and 2014, respectively. He received the Ph.D. degree in Electrical Engineering at Rensselaer Polytechnic Institute, Troy, NY, USA in 2018. He is currently an assistant professor in the School for Engineering of Matter, Transport, and Energy at Arizona State University, Tempe, AZ, USA. Before joining ASU, he was a postdoctoral researcher in the Oden Institute for Computational Engineering and Sciences at the University of Texas at Austin, Austin, TX, USA. His research interests include formal methods, autonomous systems, control systems, and reinforcement learning. He is a recipient of Howard Kaufman '62 Memorial Fellowship, ASU Fulton Schools of Engineering Top 5\% Teaching Recognition Award, and NSF CAREER Award.
\end{biography}
\vskip -2\baselineskip plus -1fil
\begin{biography}[{
\includegraphics[width=1in,height=1.25in,clip,keepaspectratio]{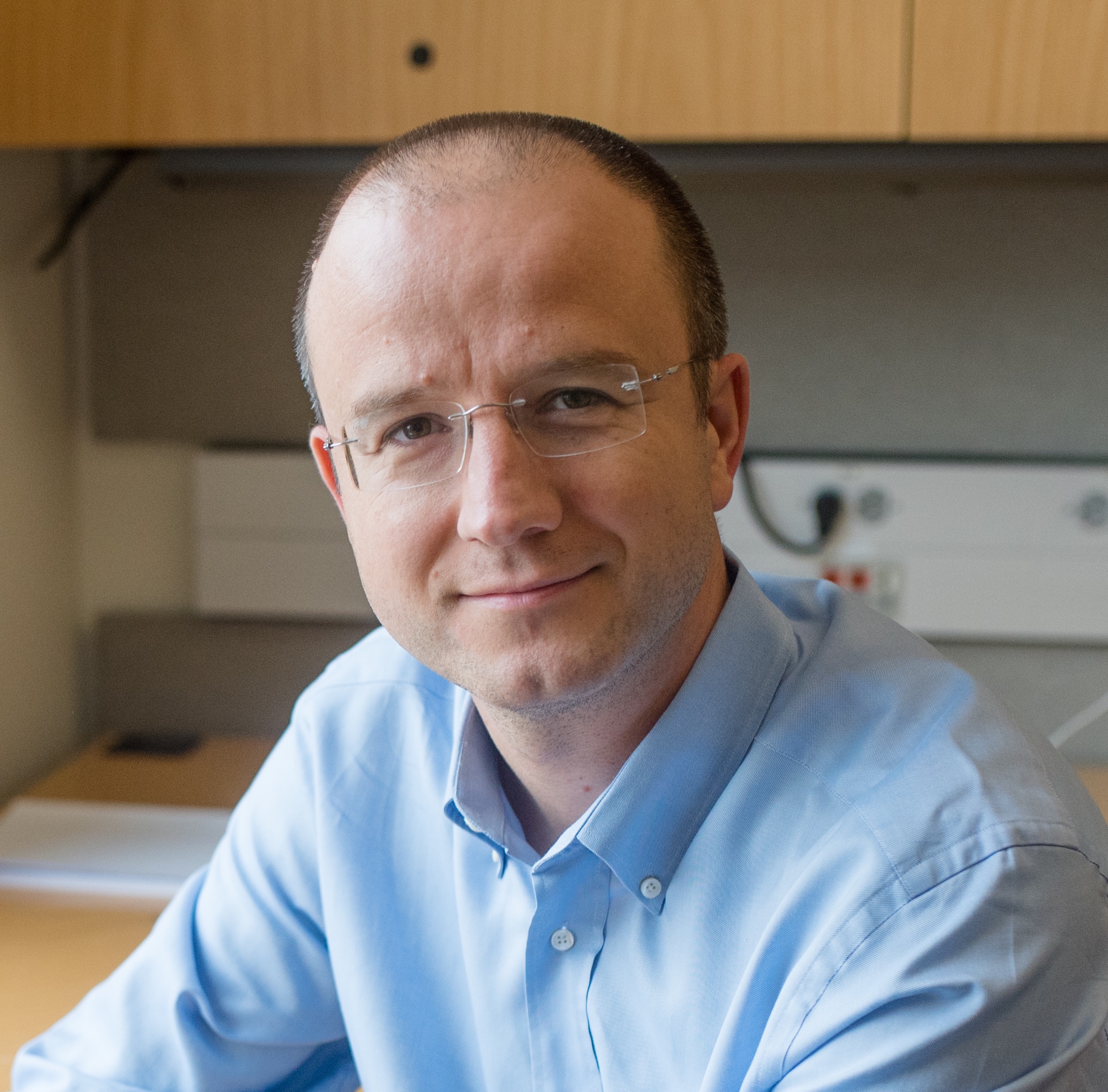}
}]{Ufuk Topcu} is a Professor in the Department of Aerospace Engineering and Engineering Mechanics at The University of Texas at Austin, where he holds the Temple Foundation Endowed Professorship No. 1 Professorship. He is a core faculty member at the Oden Institute for Computational Engineering and Sciences and Texas Robotics and the director of the Autonomous Systems Group. Ufuk’s research focuses on the theoretical and algorithmic aspects of the design and verification of autonomous systems, typically in the intersection of formal methods, reinforcement learning, and control theory.
\end{biography}
\end{document}